\documentclass[fleqn,usenatbib]{mnras}

\usepackage{newtxtext,newtxmath}

\usepackage[T1]{fontenc}
\usepackage{ae,aecompl}

\usepackage{graphicx}	
\usepackage{amsmath}	
\usepackage{amssymb}	

\title[Theoretical radial velocities of roAp stars]{A theoretical tool for the study of radial velocities in the atmospheres of roAp stars}
\author[P. Quitral-Manosalva et al.]{
Paola Quitral-Manosalva,$^{1,2}$\thanks{E-mail: Paola.Quitral@astro.ut.pt}
Margarida S. Cunha,$^{1,2}$
Oleg Kochukhov,$^{3}$
\\
$^{1}$Instituto de Astrof\'isica e Ci\^encias do Espa\c{c}o, Universidade do Porto, CAUP - Rua das Estrelas, PT4150-762 Porto, Portugal\\
$^{2}$Departamento de F\'isica e Astronomia, Faculdade de Ci\^encias, Universidade do Porto, Rua do Campo Alegre 687, PT4169-007\\ Porto, Portugal\\
$^{3}$Department of Physics and Astronomy, Uppsala University, Box 516, 751 20, Uppsala, Sweden\\
}

\date{Accepted XXX. Received YYY; in original form ZZZ}

\pubyear{2015}

\begin{document}
\label{firstpage}
\pagerange{\pageref{firstpage}--\pageref{lastpage}}
\maketitle

\begin{abstract}
Over the last decade significant amounts of high-spectral and time-resolution 
spectroscopic data have been acquired for a number of rapidly oscillating Ap stars. 
Progress in the 
understanding of the information held by these data requires the development of 
theoretical models that can be directly compared with them. In this work we present a 
theoretical model for the radial velocities of roAp stars that takes full account of the 
coupling between the pulsations and the magnetic field. We explore the impact on 
the radial velocities of changing the position of the observer, the mode frequency and 
angular degree, as well as of changing the region of the disk where the elements are 
concentrated. We find that for integrations over the full disc, in the outermost layers 
the radial velocity is generally dominated by the acoustic waves, showing a rapid 
increase in amplitude. The most significant depth-variations in the radial velocity phase 
are seen for observers directed towards the equator and for even degree modes with 
frequencies close to, or above the acoustic cutoff. Comparison between the radial 
velocities obtained for spots of elements located around the magnetic poles and around 
the magnetic equator, shows that  these present distinct amplitude-phase relations, 
resembling some of the differences seen in the observations. Finally, we discuss the 
conditions under which one may expect to find false nodes in 
the pulsation radial velocity of roAp stars.
\end{abstract}



\section{Introduction}

The rapidly oscillating Ap stars (roAp) are main-sequence classical pulsators,
with oscillations that can have periods between 6 and 24 min 
\citep{kurtz1982rapidly,alentiev2012discovery}. They are a subclass of chemically peculiar stars, 	
and have strong magnetic fields,
with mean magnitudes of a few kG \citep{mathys2017ap}.
Up to this day 61 pulsators of this type have been found \citep{smalley2015catalog}.
The pulsations are high-order p-modes that are modified in the surface layers by the magnetic field.
They are usually aligned with the magnetic field, and, in turn, inclined with respect to the rotation axis of the star,  
which makes the roAp stars oblique pulsators \citep{kurtz1982rapidly}. 

Since the first detection of radial velocity variations in roAp stars \citep{matthews1988detection},
many high-resolution spectroscopic studies
have made possible the extraction of large amounts of information
about the pulsations through the inspection of these radial velocities.
A distinguishing feature of roAp pulsations demonstrated by these studies is an unusually 
large difference in pulsation amplitudes
and phases observed in spectral lines of different chemical elements
and even different ions of the same element.
That is due to the stratification of metals, in particular rare-earth elements (REE), in the atmosphere of peculiar stars, 
which gives us the opportunity of observing different heights in the atmosphere of the star. 
Moreover, the fact that some of these elements are not uniformly distributed, but rather concentrated in spots,
means that through high-resolution spectroscopy one can probe different areas on the stellar disk. 

Through fitting the observed radial velocity to a function of the type $A \cos(\omega t + \phi)$,
where $A$ is an amplitude, $\phi$ a phase, $\omega$ the pulsation angular frequency and $t$ the time,  
these observational studies provide information on amplitude and phase variations throughout the atmospheric layers of the stars. 
Examples of this are provided in the works by \citet{kochukhov2001time, mkrtichian2003radial} and \citet{ryabchikova2007pulsation}.
In some other cases, 
radial velocity amplitude and phase shifts are derived from the bisector analysis of the spectral lines \citep{baldry1998spectroscopy,kurtz2006discovery}.

A number of different theoretical non-perturbative analyses have been developed over the years to 
address the coupling between the magnetic field and pulsations in roAp stars
\citep{dziembowski1996magnetic,bigot2000non,cunha2000magnetic,saio2004axisymmetric,cunha2006improved,
sousa2008theory,khomenko2009simulations,sousa2011understanding}. 
Among these, the models by \citet{cunha2006improved} and by \citet{saio2004axisymmetric} are particularly relevant to the current study, as they consider a realist equilibrium model, full coupling between the interior and atmosphere and allow the probing of frequencies beyond the acoustic cutoff. 
Both models show that the eigenfunctions are strongly distorted in the outer layers 
by the presence of the magnetic field,
which not only changes the amplitude of the perturbations, 
but also adds a significant angular component to the displacement.
This type of distortion has been detected also in observations by \citet{kochukhov2004indirect}.

The  theoretical models also predict shifts in the frequencies that
increase smoothly due to the effect of the magnetic field up to a point when they decrease suddenly, 
starting to increase again for frequencies still larger. 
These sudden jumps repeat periodically as the frequency increases, 
every time  the coupling between the oscillations and the magnetic field is optimal. 
In both works it was found that around these frequency jumps the eigenfunctions are most strongly perturbed, 
and their modeling becomes increasingly difficult.
In this work we use the code developed by \citet{cunha2006improved} to compute the radial velocities for roAp stars 
and compare the results with the typical amplitude and phase variations derived from observational data.
Due to the difficulty in modelling the eigenfunctions close to the frequency jumps mentioned before, 
in the present work we will not consider such frequencies.

In Section \ref{sec:model} we describe the equilibrium model of the star and the pulsation model, 
as well as the physical properties of the solutions.
We also describe the method to obtain the radial velocities in the atmosphere of the star.
In Section \ref{sec:result} we discuss seven different cases illustrating different results. 
Finally, in section  \ref{sec:discu} we discuss our results in the light of the observational data and conclude.

\section{The Model}\label{sec:model}
 
\subsection{Equilibrium model} 
 
 To model the radial velocities in the outer layers of the stars, 
 we consider small perturbations to an equilibrium model with global properties
 within the range observed for this class of pulsators
 (see in Table \ref{tab:star}). The parameters of this model also sets it within the 
 region where the excitation mechanism of roAp stars can be theoretically understood
 \citep{balmforth2001excitation,cunha2002theoretical,saio2005non,cunha2013testing}.
 
 As we are particularly interested in studying the pulsation properties in the atmosphere of the star, 
 the equilibrium model, computed with the CESAM code (Code d'Evolution Stellaire Adaptatif et Modulaire) \citep{morel1997cesam}, has had the atmosphere extended. 
In addition, we added an isothermal atmosphere on the top of the model in order to  
allow us to reach lower densities such as those found in the self consistent models of
peculiar stars' atmospheres \citep{shulyak2009model}. 
In the isothermal atmosphere the pressure and density have the form:
$p=p_s e^{-\eta/H}$ and $\rho=\rho_s e^{-\eta/H}$, respectively, were $\eta$ is the height measured from the bottom 
of the isothermal atmosphere, $p_s$ and $\rho_s$ are the pressure and density at the top of the CESAM model 
with values of $2\times10^3$ Ba and $4 \times 10^{-9}$ g cm$^{-3}$ respectively, 
and $H$ is the pressure scale height.

The atmospheric structure of roAp stars is very complex, showing horizontal and vertical variations of chemical elements and possible gradients of the magnetic field intensity. These properties have been studied in a number of works \citep{nesvacil2004probable,shulyak2009model,kochukhov2009self,shulyak2010realistic,kudryavtsev2012magnetic,nesvacil2013self}.
In general, these studies showed that, although the atmospheric structure deviates systematically from that of normal stars, it is 
not particularly anomalous in the sense that it can still be approximated by a steep 
temperature decline followed by roughly isothermal upper layers.
That justifies the model 
adopted here. The exception is a small temperature inversion associated with the 
high-lying REE cloud (see illustrations in
\citet{shulyak2009model,nesvacil2013self}),
which we have not considered in our model and that may have implications to the 
reflection of the waves. This will be referred in section \ref{sec:discuobs}.

 Furthermore, we assume that the magnetic field is force-free, 
 and neglect the effect of rotation both on the equilibrium structure and on the pulsations. 
 In practice we will consider in this work a dipolar magnetic field with a polar magnitude $B_p$.
 
\begin{figure}
	\includegraphics[clip,width=\columnwidth]{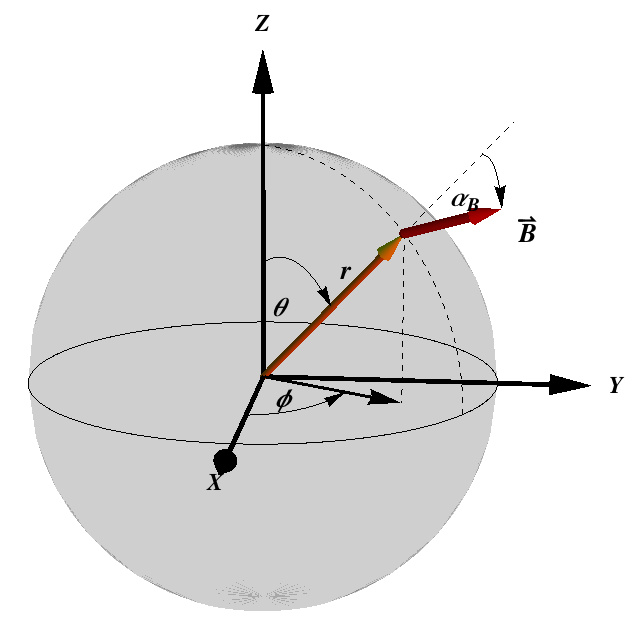}
    \caption{ Illustration of the star, 
    permeated by a dipolar magnetic field with identification of the  Cartesian coordinates ($X,Y,Z$), 
    and the spherical coordinates ($r,\theta,\phi$).
    Shown is also the vector magnetic field, $\vec{B}$, at a co-latitude $\theta$ and the angle $\alpha_B$ 
    between the radial direction and the direction of the local magnetic field.
    }
    \label{fig:Coordenadas}
\end{figure}

 \begin{table}
	\centering
	\caption{Parameters of the stellar model considered in this work. Mass, radius, effective temperature, temperature of the isothermal atmosphere, and 
	acoustic cut-off frequency.  }
	\label{tab:star}
	\begin{tabular}{lcccr} 
		\hline
		Mass & Radius & $T_{\mathrm{eff}}$ & $T_{\mathrm{iso}}$ & $f_{\mathrm{cut-off}}$ \\
		\hline
		$1.8 $M$_{\odot}$ & $1.57$R$_{\odot}$ & $8363$K & $6822$K & $2.458$mHz\\
		\hline
	\end{tabular}
\end{table}
In Fig. \ref{fig:Coordenadas} we illustrate the magnetic field in the equilibrium.
The magnetic field axis is along the Z direction in the Cartesian 
coordinate system $(X,Y,Z)$, while $(r,\theta,\phi)$ are the spherical coordinates. 
We show the magnetic field $\vec{B}$ at the co-latitude $\theta$ and define
the angle between the position vector $\vec{r}$ and the magnetic field $\vec{B}$ as $\alpha_B$.   
 
\subsection{Pulsation model} 

The pulsation model is based on that adopted for the MAPPA code (MAgnetic Perturbations to Pulsations in Ap stars) \citep{cunha2006improved}.
In that model two regions are considered, namely, the interior of the star
where the magnetic pressure is neglected and the outer layers, named by the author the magnetic boundary layer,
where the magnetic pressure is comparable or larger than the gas pressure.
In the interior, the standard oscillation model is used to describe the p-modes.
In the Magnetic boundary layer, with a characteristic depth of only a few \% of the radius,
the author considers the direct effect of the magnetic field on the pulsations, and describes them by
the following system of magnetohydrodynamic equations: 

\begin{equation}
\dfrac{\partial \vec{B}}{\partial t}= \triangledown \times (\vec{v}\times \vec{B} ),
\label{eq:1system}
\end{equation}
 
\begin{equation}
\dfrac{D\rho}{D t}+\rho \triangledown \cdot \vec{v} = 0,
\label{eq:2system}
\end{equation}

\begin{equation}
\rho \dfrac{D\vec{v}}{D t} = - \triangledown p +\vec{j}\times \vec{B} + \rho \vec{g},
\label{eq:3system}
\end{equation}

\begin{equation}
\dfrac{D p}{D t} = \frac{\gamma p}{\rho} \dfrac{D \rho}{D t},
\label{eq:4system}
\end{equation}
where the current density is $\vec{j}= 1/\mu_0 \, \nabla \times \vec{B}$,
$\mu_0$ is the permeability in the vacuum, $\rho$ is the density of the gas, $p$ is the pressure, 
$\vec{g}$ is the gravitational field, $\vec{\xi}$ is the displacement vector and $\vec{v}=\partial\vec{\xi}/\partial t$
is the velocity. 
The system represents adiabatic motions,
in the limit of perfect conductivity
and is solved for small perturbations to the equilibrium structure and under the Cowling approximation.

Since the magnetic boundary layer is thin and the magnetic field varies on large scales only, 
the equations in this region of the star are solved by performing a plane-parallel approximation 
and assuming a local constant magnetic field, at each latitude. Consequently,
at each latitude a local-coordinate system $(x,y,z)$ is defined with 
the $z$ component pointing outwards of the star, 
and such that the magnetic field is zero in the $y$ direction.   
The local magnetic field at a given co-latitude $\theta$, is then given by,
\begin{equation}
\vec{B}= \left( \frac{B_p}{2}\sin(\theta), 0,B_p\cos(\theta)\right),
\end{equation}
where $B_p$ is assumed to be constant, which is a good approximation given that the layer is thin.

Furthermore, since the system is solved under a linear approximation
it does not inform about the amplitude of the displacement. 

In what follows we will consider only solutions corresponding to the azimuthal order $m=0$,
thus, the solution for the displacement at each latitude, 
in the local coordinate system, will be written as $\vec{\xi}=(\xi_x,0,\xi_z)$.
Moreover, a second coordinate system will be used in the local approximation, 
namely, one that has axes parallel and perpendicular to the magnetic field. 
The latter coordinate system is obtained from the first through a rotation of $\alpha_B$
around the $y$ axis. We denote it by $(u_{\parallel},y,u_{\perp})$. 
In the second coordinate system the solutions are written in the form $\vec{\xi}=(\xi_{\parallel},0,\xi_{\bot})$.

In order to understand the solutions given by the MAPPA code \citep{cunha2006improved} 
we need to consider separately the two different regions mentioned before, namely 
the magnetic boundary layer of the star, and the interior.
In the latter, dominated by the pressure of the gas, we find two decoupled solutions, an acoustic  wave,
that is displacing the gas in the radial direction, 
and a transverse Alfv\'en wave that is displacing the gas in a local horizontal direction \citep{cunha2000magnetic,dziembowski1996magnetic}.
Moreover, in the magnetic boundary layer the solutions can be best understood by further dividing this layer into  two different regions \citep{cunha2007theory}, namely, 
the region where the pressure of the gas is of the same order of magnitude as the magnetic pressure 
and the outermost layers, where the magnetic pressure dominates.  
In the former, we have magnetoacoustic waves,
while in the latter the waves decouple once again in the form of acoustic waves that are displacing the gas in the direction 
parallel of the magnetic field, and of compressional  Alfv\'en waves,
that are displacing the gas in the direction perpendicular to the magnetic field \citep{sousa2008mode}.

\subsection{Decoupling of the waves}

The system of equation (\ref{eq:1system})-(\ref{eq:4system}) is solved up 
to a normalizing constant by applying adequate boundary conditions.
At the surface the magnetic field is matched continuously onto a vacuum field.
The remaining boundary conditions are obtained by matching the numerical solutions to 
approximate analytical solutions in the regions where the magnetic and acoustic waves are decoupled,
as described below. 

In the interior of the star the acoustic component corresponds to the solution obtained when $\vec{B}=0$ and 
the magnetic component is assumed to be a wave that dissipates inside the star. 
Then, there the numerical solution for the magnetic component is matched onto an 
analytical asymptotic solution for an Alfv\'en wave propagating 
towards the interior of the star (see \citet{cunha2000magnetic} for details). 

The final boundary condition consists in matching the numerical solution for the parallel component of the displacement 
to its analytical counterpart in the isothermal atmosphere.
In the isothermal atmosphere the analytical solutions are those derived in the work of \citet{sousa2011understanding}. 
There the magnetoacoustic waves are already decoupled into a (slow) acoustic wave and a (fast) compressional Alfv\'en wave
that move in the directions parallel and perpendicular to the magnetic field, respectively, and have the form, 
\begin{equation}
\xi_{\parallel}= \frac{|A_s|}{p^{1/2}} \exp{i(\pm k_{\parallel} \eta+\omega t+\phi_s )},
\label{eq.sound_iso}
\end{equation}
\begin{equation}
\xi_{\bot}= |A_f| J_0(2\sqrt{\chi \rho} ) \exp{i(\omega t+\phi_f )},
\label{eq.mag_iso}
\end{equation}
where $\omega$ is the angular oscillation frequency, $A_s$,$\phi_s$,$A_f$,$\phi_f$ are the amplitudes (A) 
and phases ($\phi$) of the acoustic and magnetic waves respectively,
at the bottom of the isothermal atmosphere, that depend on the latitude, 
$J_0$ is the Bessel function and $\chi$ is a constant $\chi=H^2 \omega^2 \mu_0 / B_p^2$.
Moreover, the parallel component of the wavenumber is defined by:
\begin{equation}
k_{\parallel}= \sqrt{\frac{\omega^2 \rho}{\gamma p \cos^2 \alpha_B}-\frac{1}{4H^2}},
\label{eq.k_pa}
\end{equation} 
where $\gamma$ is the fist adiabatic exponent.
$k_{\parallel}$ can be real or imaginary. In the former case the parallel component of the solution (acoustic wave) is oscillatory,
while in the latter case it is exponential.

Inspecting the parallel component of the wave number $k_{\parallel}$, 
we can see that it depends on latitude through the angle $\alpha_B$.
Therefore, even when the frequency of the oscillation is below the acoustic cut-off frequency
for a non-magnetic star, in the presence of a magnetic field  
$k_{\parallel}$ will become real and the solutions will become oscillatory
when the co-latitude is larger than a given critical value.
The critical frequency,
\begin{equation}
\omega_c=\sqrt{\frac{\gamma p \cos^2 \alpha_B}{4H^2\rho}},
\label{eq.wc}
\end{equation} 
defines the co-latitude at which the parallel component of the solution changes its behavior
from exponential to oscillatory in the presence of the magnetic field. 
We shall call that co-latitude the critical angle, $\alpha_{cr}$

For a dipolar magnetic field, the critical frequency has its maximum value at the magnetic pole,
corresponding to the the critical frequency in the absence of a magnetic field,
(i.e. to the acoustic cut-off frequency).
But it decreases as the magnetic equator is approached.
As a consequence, even if the oscillation frequency is below the acoustic cut-off frequency, 
it will always be above the local critical frequency for co-latitudes larger than a critical value and,
thus,
there will always going to be wave energy losses in the equatorial zone when the magnetic field is considered.

\subsection{Displacement solutions}
\label{subsec:solut} 

 Due to the structure of the magnetic field, at each latitude the distortion of the oscillation is different. 
 In particular, the magnitude and direction of the magnetic field is different at different latitudes, 
 affecting differently both the amplitude and characteristic scale of the displacement.
 To illustrate this, we discuss below a particular case, at two different latitudes.
 
\begin{figure}
	\includegraphics[width=\columnwidth]{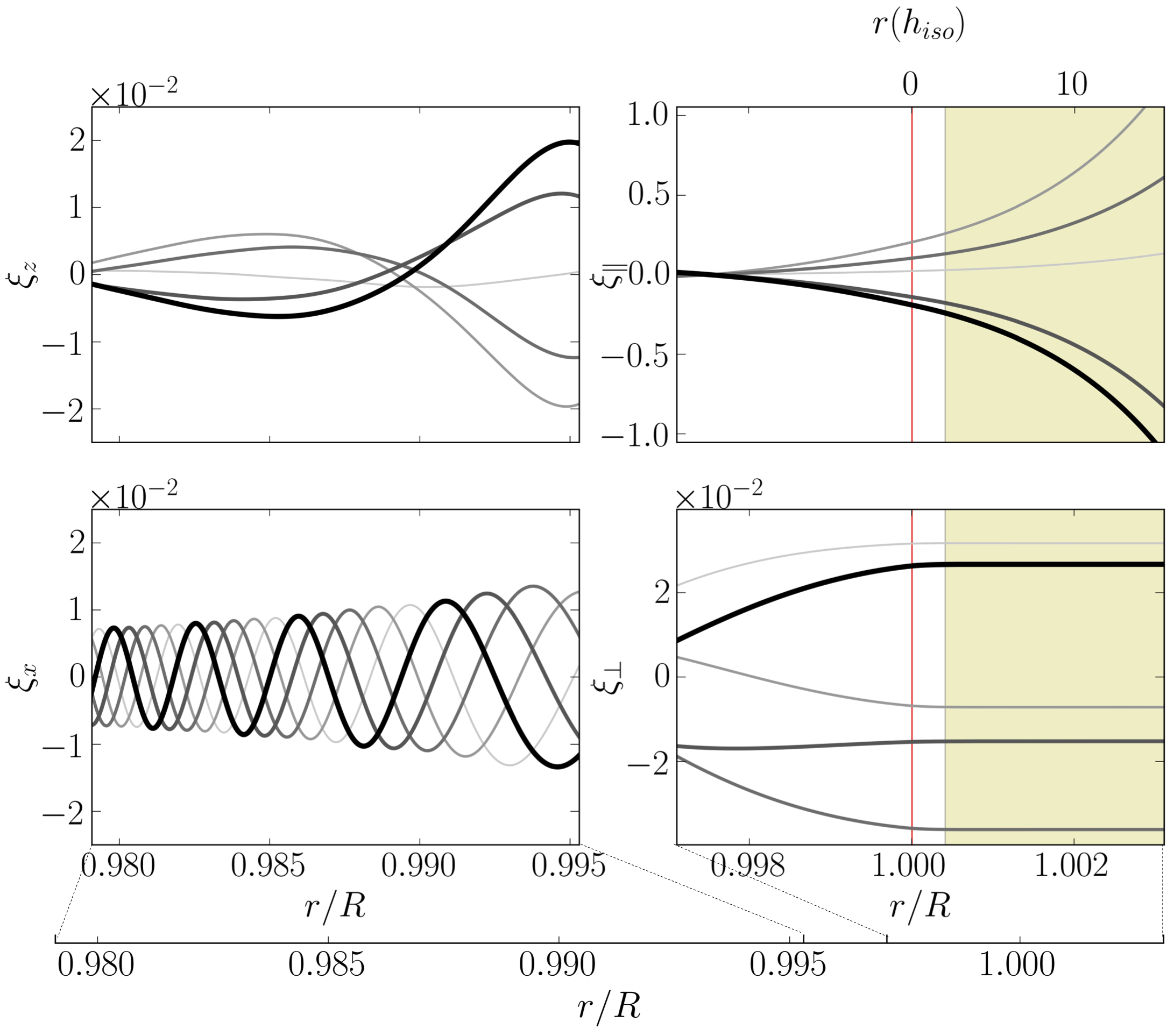}
    \caption{Dimensionless displacement $\vec{\xi}$ at the co-latitude of $37^\circ$, as a function of the normalized radius (bottom x axis) 
    in the outer $2\%$ of the stellar model,
    for a frequency $1.7$ mHz with a magnetic field of $2$ kG.
    The left panels show the components of the displacement in the local vertical (top) 
    and local horizontal (bottom) directions. 
    The right panels show the displacement in the direction parallel to the magnetic field (top),
    and perpendicular to the magnetic field (bottom).
    The different curves represent different times, the yellow shadow marks the isothermal atmosphere, 
    and the red vertical line represents the bottom of the photosphere of the star. 
    The top x axis indicates the atmospheric  height measured from the bottom of the photosphere in units of the (constant) pressure scale height of the isothermal atmosphere. 
    }
    \label{fig:Xi_mayor}
\end{figure}

\begin{figure}
	\includegraphics[width=\columnwidth]{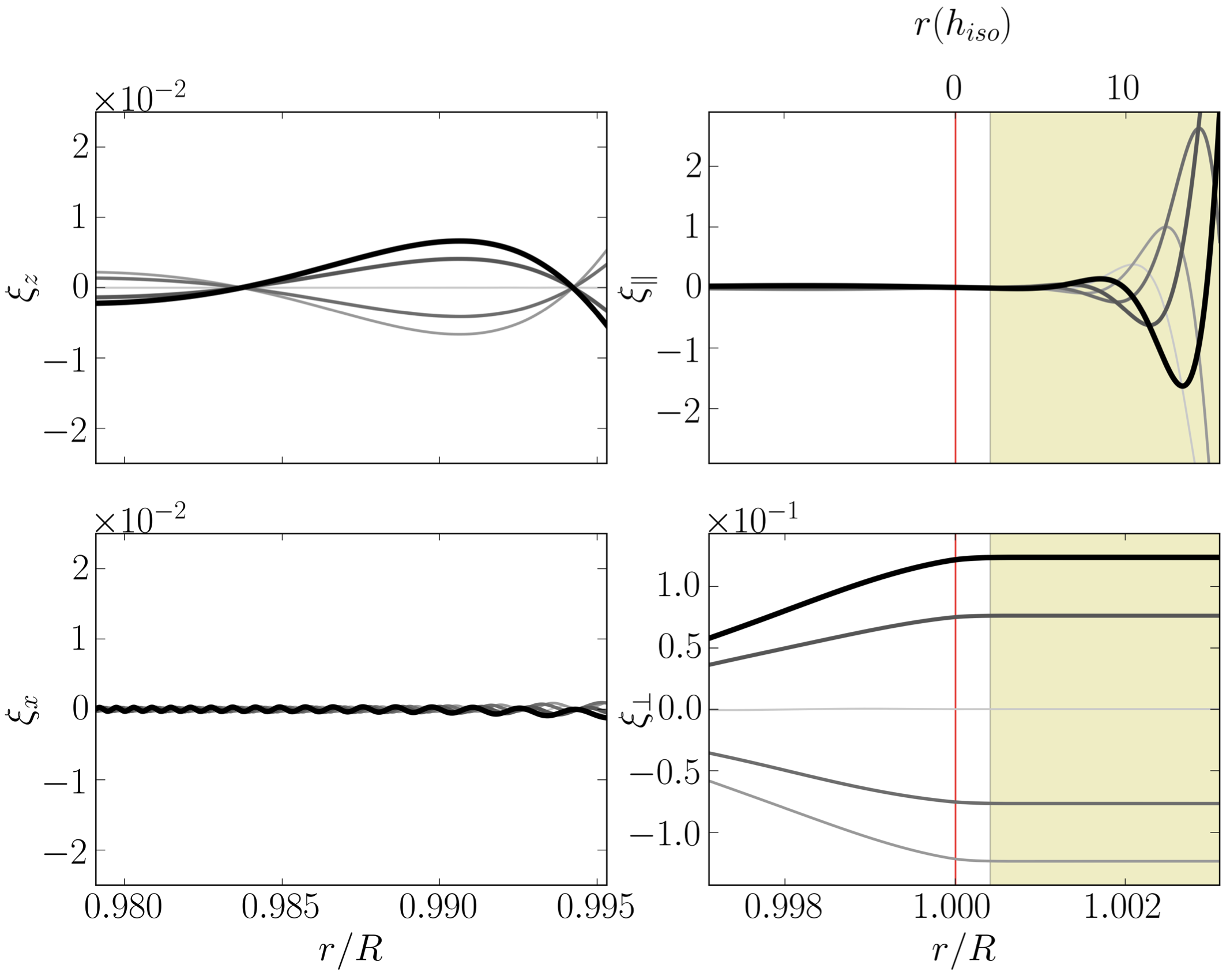}
    \caption{The same as in figure \ref{fig:Xi_mayor} but for a co-latitude of $87^\circ$.}
    \label{fig:Xi_menor}
\end{figure}

 The displacement as a function of the radius at two different latitudes is shown for a cyclic frequency ($f =\omega/2\pi$) of $1.7$ mHz 
 and a magnetic field, $B_p$, of $2$ kG, in Figs. \ref{fig:Xi_mayor} and \ref{fig:Xi_menor}.
 On the left panels we show the components of the solution in the innermost part of the magnetic boundary layer, 
 using the local coordinate system (x,y,z), 
 and on the right panels the local parallel and perpendicular components of the solution in the outermost part of the magnetic boundary layer, using the coordinate system $(u_{\parallel},y,u_{\perp})$.
 
 In Fig. \ref{fig:Xi_mayor} the displacement is shown for a co-latitude of $37^\circ$. 
 At this latitude the frequency is below the critical frequency and, 
 thus, in the isothermal atmosphere, marked by the yellow-shaded region on the right panels,
 the acoustic wave (component of the displacement parallel to the magnetic field) shows a standing behavior 
 as does the compressional Alfv\'en wave (perpendicular component), which shows, 
 in addition, a constant amplitude in that part of the atmosphere, as expected from eq. (\ref{eq.mag_iso}).
 
 In the inner layers shown on the left panels, the acoustic wave (vertical component $\xi_z$ in these layers), 
 also presents an almost standing behavior while the Alfv\'en wave (the horizontal component $\xi_x$) 
 has a clear running behavior, dissipating towards the interior of the star as expected from the boundary conditions.
 
 Figure \ref{fig:Xi_menor} illustrates a case of a co-latitude of $87^\circ$.  
 As expected, closer to the equator, 
 where the frequency of the wave is larger than the critical frequency,
 the acoustic wave will, instead, have a running behavior in the atmosphere.
 As a consequence, 
 at this co-latitude the exponential growth of the wave amplitude is larger than at the co-latitude of $37^\circ$.
 Because the energy carried by the acoustic wave is conserved, 
 the wave amplitude thus increases as the inverse of the root square of the density 
 (or, equivalently in these layers, of the root square of the pressure- cf eq. (\ref{eq.sound_iso})). 
 We note, for comparison,  that for the co-latitude of $37^\circ$, the exponential term in eq. (\ref{eq.sound_iso}) 
 partially compensates the exponential increase associated with the decrease of the pressure,
 leading to a smaller increasing rate of the amplitude, consistent with a decrease in the energy content of the acoustic wave.

\subsection{Radial velocities}

To obtain the radial velocity as a function of the radius in the outer layers of a roAp star
we need to integrate the velocity field at each specific radius over the area of interest,
which may be the full visible disc, or a sub-section of it, 
when the elements contributing to the radial velocity measurement are concentrated in a particular region only.  
We compute the integrated velocity field, considering a liner limb-darkening law, using the expression 
\citep{dziembowski1977light},
\begin{equation}
 V_{int}=\int\limits_{\varphi'_i}^{\varphi'_f} \int\limits_{\theta'_i}^{\theta'_f}
\left[ v_{r} X_{r}+v_{\theta} X_{\theta}\right]  
\times C_n^{-1} (1-a(1-\cos{\theta'})) \cos{\theta'} \sin{\theta'} d\theta' d\varphi',
 \label{eq:vh}
\end{equation} 
where $(r,\theta,\phi)$ is the spherical coordinate system, described in Fig. \ref{fig:Coordenadas},
$(r,\theta',\phi')$ is a spherical coordinate system with the polar axis, 
here named $Z^{\prime}$, aligned with the direction of the observer,
$a$ is the limb-darkening coefficient, for which we adopt a value of 0.46 \citep{claret2003limb},
and $C_n$ is a normalization constant from the integration of the limb-darkening
in the visible disc.
$\varphi'_i$, $\varphi'_f$ and $\theta'_i$, $\theta'_f$, are the integration limits, that represent a given area of the visible disk. 
$X_{r}$ and $X_{\theta}$ are the projections of the unit vector along the radial direction $\widehat{r}$ 
and along the polar direction, $\widehat{\theta}$ onto the direction of the observer $\widehat{z}'$, respectively. 
Moreover, $v_r $ and $v_{\theta}$ are the velocity components derived from the displacement, 
$\vec{v}=\dfrac{d}{dt}\delta \vec{r}$, where,
\begin{equation}
\delta \vec{r} = \left(  \xi_r(r,\theta) Y_l^0\,\, \widehat{r} + \xi_{\theta}(r,\theta) Y_l^0 \,\,\widehat{\theta} \right) \emph{e}^{\textit{i} \omega t}.
\label{eq:des}
\end{equation} 
Here $\xi_r$ and $\xi_{\theta}$ are obtained by combining the local solutions $\xi_z$ and $\xi_x$, respectively, at each latitude. 
Their $\theta$ dependence is a consequence of the presence of the magnetic field which, as discussed before, 
influences the eigenfunction  differently at different latitudes distorting the eigenfunctions from the pure spherical harmonic solutions obtained in the non-magnetic case. 
Moreover, since the system loses energy both from the running magnetic waves at the bottom of the magnetic boundary layer and from the acoustic running waves in the atmosphere, 
the eigenfrequencies and eigenfunctions are complex.
   
\begin{figure}
	\includegraphics[width=\columnwidth]{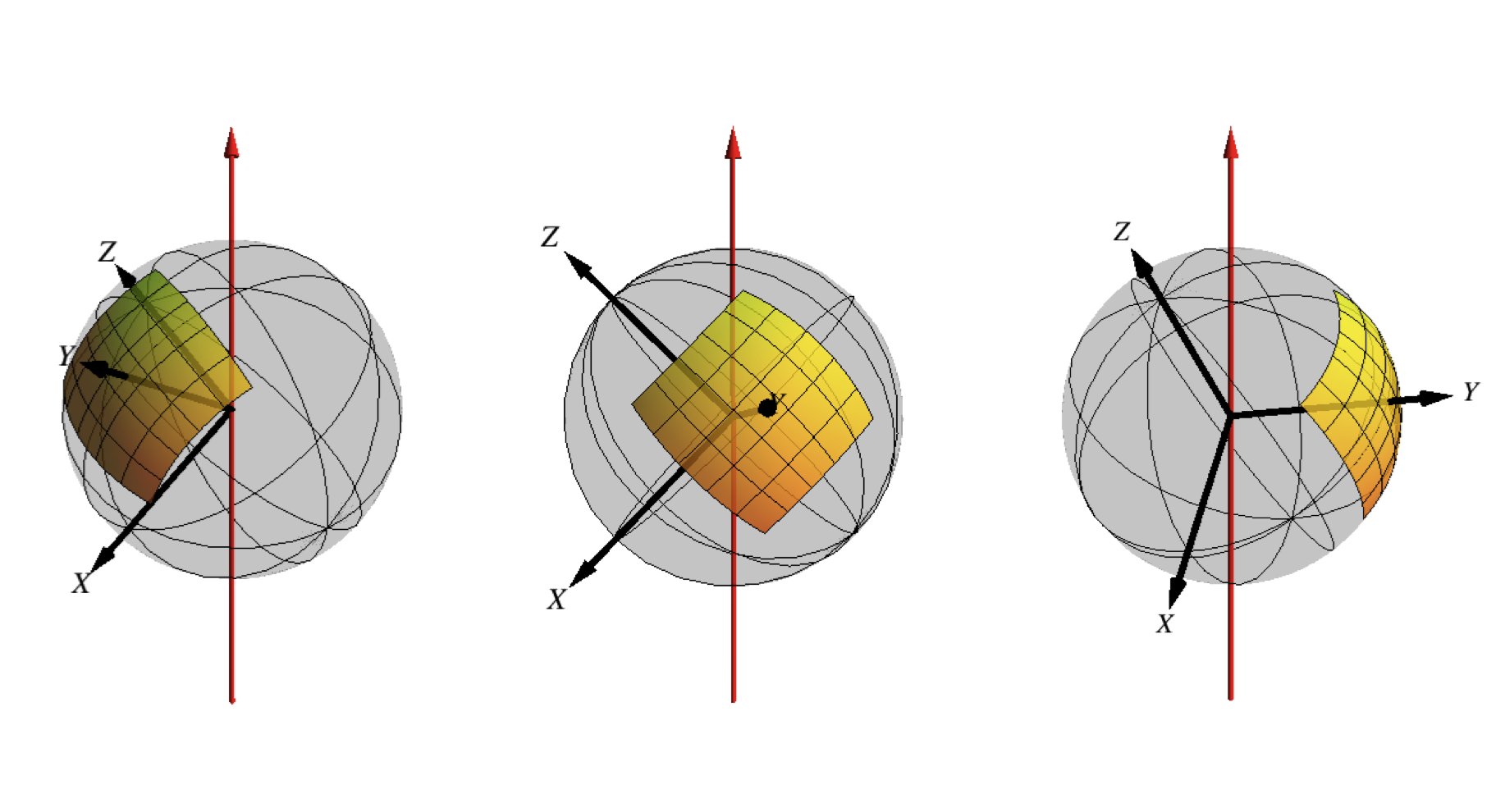}   
    \caption{Representation of a spot in the code to calculate the radial velocity,
    considering the star at different rotational phases.}
    \label{fig:roration}
\end{figure}

When considering the integration only in a certain region of the disk, 
corresponding to a chemical overabundance spot,
the spot can be studied from different viewing angles, as illustrated in Fig. \ref{fig:roration}. 
We can, thus, study the changes in the radial velocity associated to a spot throughout the rotation of the star.

We also computed the integral for the radial velocity considering, instead, 
the limb-darkening and line-weighting proposed by \citep{landstreet2000magnetic}. 
However, using their main values for the coefficients proposed, we did not find a significant difference
when comparing to the results obtained with eq (\ref{eq:vh}).
  
To facilitate the physical interpretation of the radial velocity $V_{int}$, 
 the code allows us to separate the contributions to $V_{int}$ of the different components of the velocity perturbation. 
 This is done by performing an integral that is in all equal to that defined in Eq. (\ref{eq:vh}), 
 except that the total velocity projected in the direction of the observer 
 that enters that integral is substituted by the projection of a single component of the velocity. 
 This facilitates the physical interpretation because in the outer atmospheric layers 
 the acoustic waves correspond to displacements parallel to the magnetic field
 while the magnetic waves correspond to displacements perpendicular to the magnetic field. 
 When the velocity field component aligned with the magnetic field 
 alone is considered in that integral, 
 we denominate the result of that integral $V_{\parallel}$. 
 Similarly,
 when the integral is based on the velocity field component 
 perpendicular to the magnetic field alone, 
 we denominate the result of the integral by $V_{\perp}$.
 Moreover, we compute similar integrals considering the radial and polar components of the velocity denominating the results by $V_{r}$ and $V_{\theta}$, respectively.

In summary, the code allows us to compute the radial velocity associated to the stellar pulsations, 
either for the full visible disk or for part of it.
Being able to define any area in the surface of the sphere,
that can represent a spot or a belt of elements in the atmosphere of the star, 
and redefining the limits of integration, $\varphi'_i$, $\varphi'_f$, $\theta'_i$ and $\theta'_f$, 
so that these always remain in the visible disk, the code makes it
possible to study the pulsations for different positions of the observer, 
for a single spot or the full visible disk.
In addition, it allows us to study the contributions to the radial velocity of the different components of the velocity field.

\section{Results} \label{sec:result}

 To analyze different possible solutions, we fix the magnetic field in $2$ kG and explore 3 different pulsation frequencies.
 For the first three cases we consider that the observer is pole-on and that the mode degree is $l=1$.
 In addition, we analyze three cases in which the observer's position is equator-on and the mode degree is $l=0$.
 In doing so, in particular by fixing the frequencies, we intentionally ignore the difference in frequency that would result from solving the eigenvalue problem for modes of different degrees.
We note that we did not consider an odd degree mode for the equator-on view 
because of the strong cancellation effect that would be present
 when performing the disk integration. 
 These cases correspond to the first 6 entries in Table \ref{tab:cases}.
 
   \begin{table}
	\centering
	\caption{Properties and parameters of the cases explored in this work.
	The columns are:
	frequency, $f$; polar magnetic field, $B_p$; observer's view;
	integration area, that can be of the full visible disk (F. V. D.), or of a belt in the equatorial zone (E. Z.);
	and the critical angle, $\alpha_{cr}$.  
	}
	\label{tab:cases}
	\begin{tabular}{lcccccc} 
		\hline
		     & $f$  & $B_p$ &  Obs.  & l & Int. & $\alpha_{cr}$ \\
		     & mHz  &  kG   &  from  &   & area &              \\
		\hline
		Case 1 & 1.7 & 2.0 & pole    & 1 & F. V. D. & $50^\circ$ \\
		Case 2 & 2.2 & 2.0 & pole    & 1 & F. V. D. & $33^\circ$ \\
		Case 3 & 2.7 & 2.0 & pole    & 1 & F. V. D. & $0^\circ$  \\
		Case 4 & 1.7 & 2.0 & equator & 0 & F. V. D. & $50^\circ$ \\
		Case 5 & 2.2 & 2.0 & equator & 0 & F. V. D. & $33^\circ$ \\
		Case 6 & 2.7 & 2.0 & equator & 0 & F. V. D. & $0^\circ$  \\
		Case 7 & 2.7 & 2.0 & pole    & 0 &   E. Z.  & $0^\circ$ \\
		\hline
	\end{tabular}
  \end{table}
 
\begin{figure}
	\includegraphics[width=0.9\columnwidth]{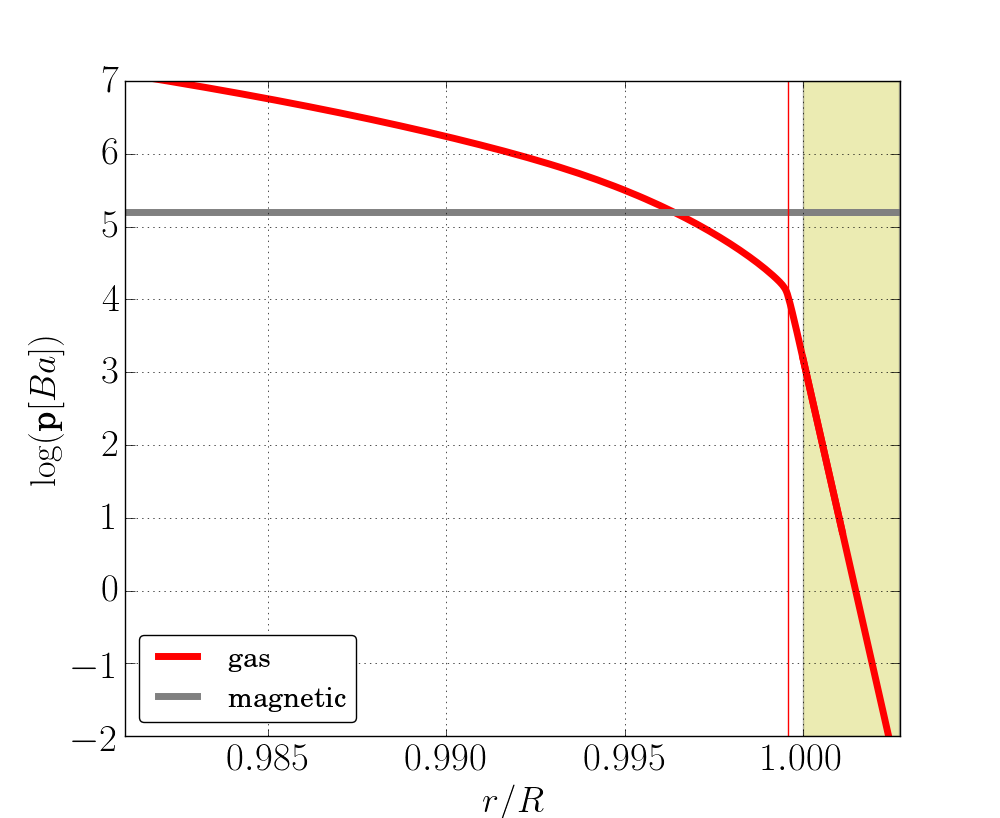}   
    \caption{Comparison between the gas pressure (thick red line) and the magnetic pressure (grey line)
    in the outer 2\% of the star for a magnetic field of 2 kG. 
    The thin vertical red line marks the bottom of the photosphere.}
    \label{fig:pres}
\end{figure}

 For a $2$ kG magnetic field, the magnetoacoustic region in our stellar model is placed fully in the interior of the star. 
 This is illustrated in Fig. \ref{fig:pres} where the gas and magnetic pressures are compared.   
 This means that in the atmospheric region the acoustic and magnetic waves are completely decoupled
 and, as noted in the section \ref{sec:model}, the acoustic waves move in the direction of the local magnetic field,
 and the magnetic waves move perpendicularly to it.
 
In addition, we consider a case in which an apparent node in the isothermal atmosphere can be seen. This corresponds to the last entry in Table \ref{tab:cases}.

 To compare the amplitude, $A_r$, and phase, $\phi_r$, variations of the theoretical radial velocity 
 with those derived from observations (e.g. \citet{ryabchikova2007pulsation}),
 we match the numerical solutions in the atmosphere to a function of the type,
 \begin{equation}
 V_{int}=A_r \cos(\omega t + \phi_r).
 \label{eq:am_pha}
 \end{equation}

\subsection{Case 1}

\begin{figure}
	\includegraphics[width=\columnwidth]{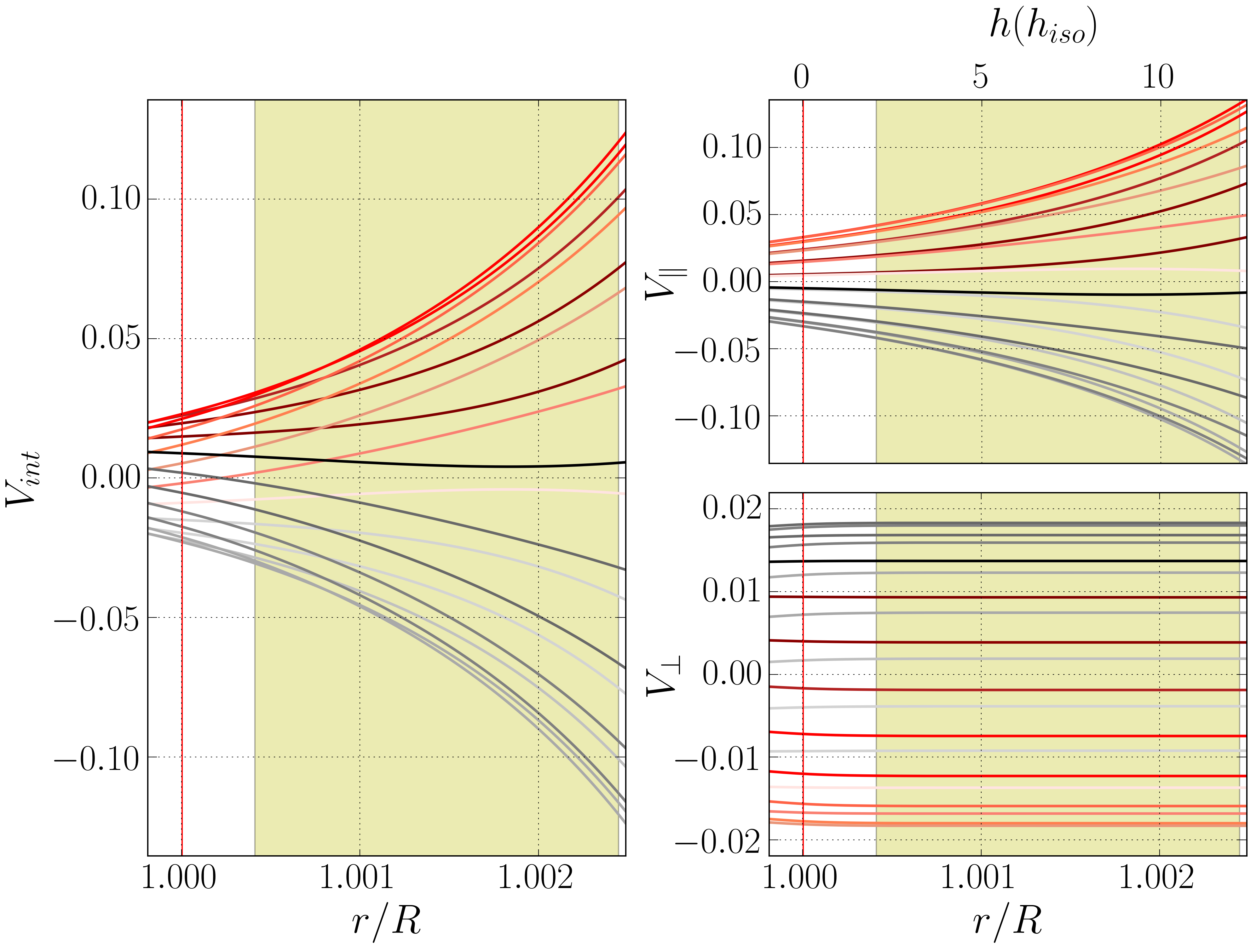}
    \caption{Dimensionless and normalized radial velocity. 
    This case is for integration over the visible disc, a magnetic field of $2$~kG,
    a mode of frequency $1.7$~mHz and degree $l=1$, and an observer pole-on.
    Shown in the left panel is the radial velocity as a function of the radius, 
    at different times within the oscillatory period
    represented by curves of different colors.   
    The upper right panel shows the contribution to the radial velocity 
    of the velocity component parallel to the magnetic field.
    The bottom right panel shows the contribution to the radial velocity of the velocity 
    component perpendicular to the magnetic field.
    The labels on the top horizontal axes show the height measured from the bottom of the photosphere
    in units of the (constant) pressure scale height of the isothermal atmosphere. 
    The red vertical line represents the bottom of the photosphere and the yellow shadow
    region represents the isothermal atmosphere.    
    }
    \label{fig:intRV_1}
\end{figure}

 The first case we discuss, is one in which the phase is found to be nearly constant.
 To find a solution with a constant phase it is necessary to choose a frequency well below the acoustic cut-off.
 Here we take a frequency of $1.7$ mHz. 
 As we mentioned earlier, we fix  the magnetic field at $B_p=2$~kG, and consider a mode of degree $l=1$ with an observer pole-on.
 At this particular frequency, the acoustic waves change from having a standing character 
 to having a running character at a critical angle $\alpha_{cr}=50^{\circ}$ (see Table \ref{tab:cases}), 
 meaning that for co-latitudes larger than this angle the local critical frequency is smaller than $1.7$ mHz.
 The radial velocity is shown in the left panel of Fig. \ref{fig:intRV_1}.  
 The contributions of the components of the velocity parallel and perpendicular to the magnetic field to the integral that defines 
 the radial velocity (eq. (\ref{eq:vh})) are presented in the top- and bottom-right panels, respectively.
 In each panel the vertical red line indicates the bottom of the photosphere of the star 
 and the shaded-yellow area represents the isothermal atmosphere. 
 The same notation is used for all other cases.

Looking at the contribution of the parallel and perpendicular velocity components in the 
atmosphere of the star, we can verify that the acoustic and magnetic waves are already decoupled, 
since, as predicted analytically by eqs. (\ref{eq.sound_iso})-(\ref{eq.mag_iso}),
we see an exponential behavior for the parallel component, and a constant behavior for the 
perpendicular component.
Their contribution to the radial velocity integral is of similar magnitude, although the acoustic waves become progressively dominant with increasing atmospheric height.
      
\begin{figure}
	\includegraphics[width=4.01cm]{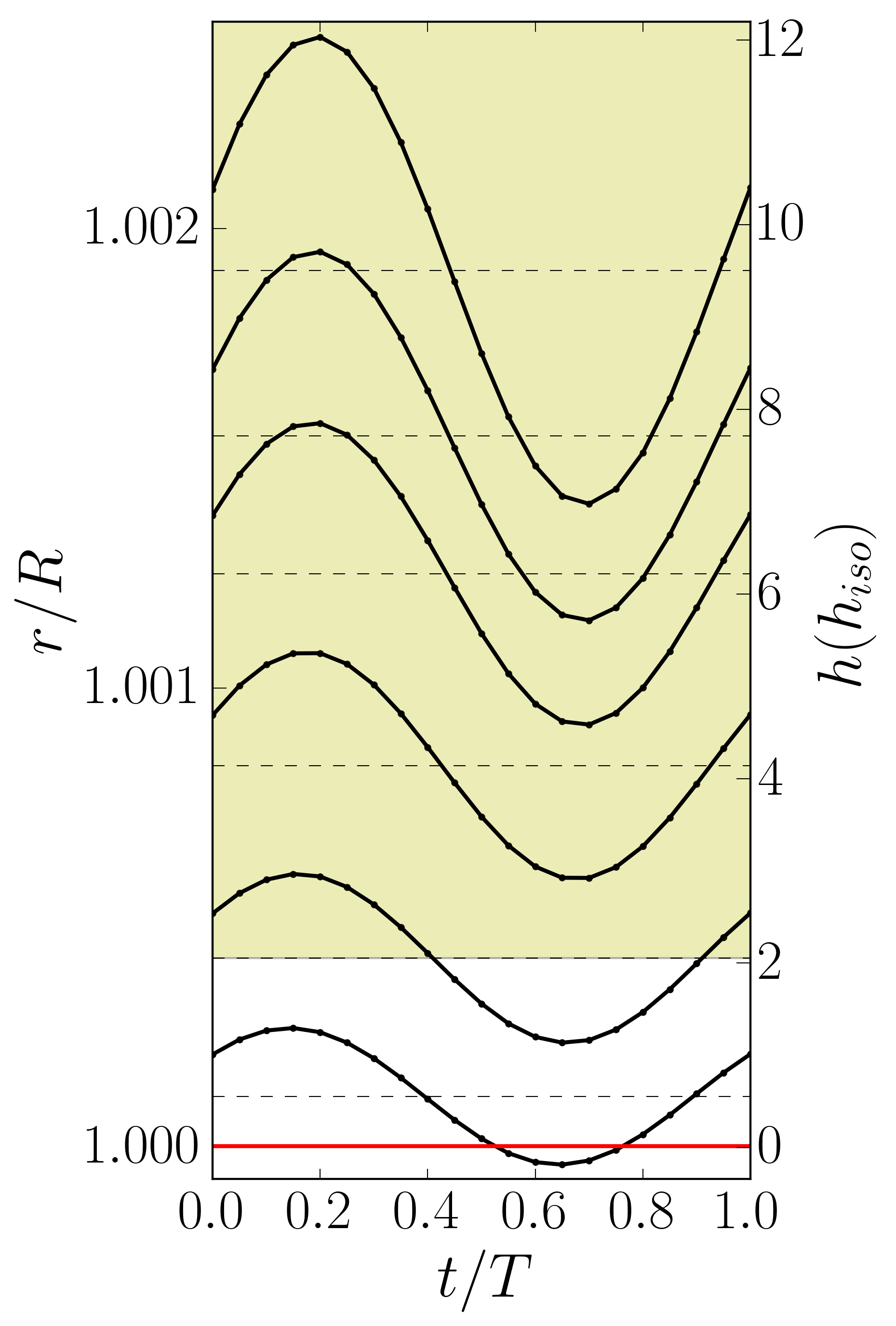}
    \includegraphics[width=4.2cm]{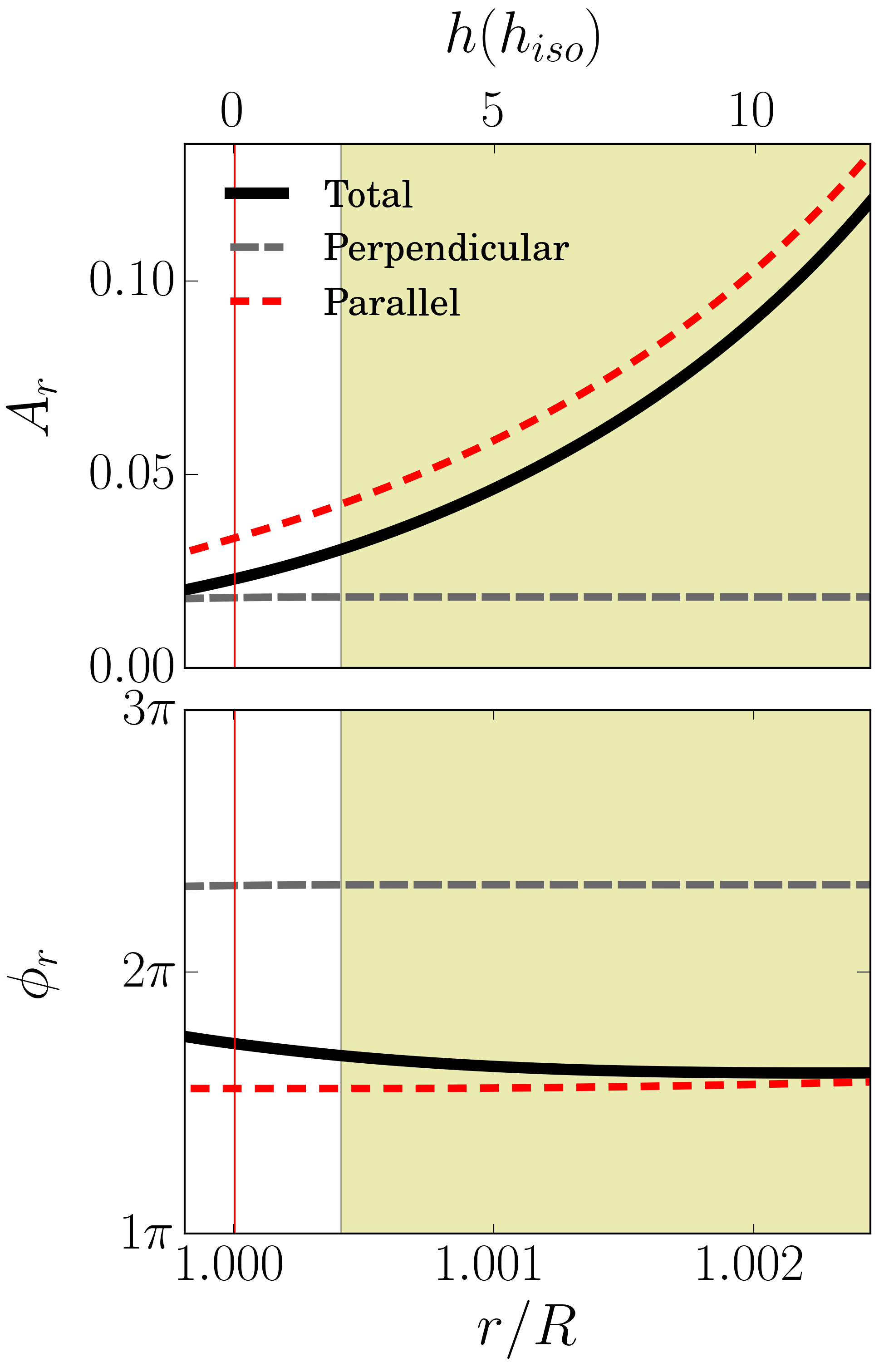}
	\caption{The left panel shows the amplitude of the radial velocity at different heights in the atmosphere, 
	as a function of the fraction of time within an oscillation period T.  
    The right panels show the amplitude (top) and the phase (bottom) 
    of the radial velocity fitted to the function given by eq. (\ref{eq:am_pha}).
    The radial velocity amplitude and phase are in black, 
    the amplitude and phase derived from $V_{\parallel}$ in red, and
    the amplitude and phase derived from $V_{\perp}$ in grey.
    The horizontal red line on the left panel and the vertical red line
    on the right panel mark the bottom of the photosphere.
    }
    \label{fig:atm_Am_ph_1}
\end{figure}

 The amplitude and phase of the radial velocity for this case are shown in Fig. \ref{fig:atm_Am_ph_1}.
 The left panel shows the amplitude variation during one period of the oscillation at different heights 
 in the atmosphere. We recall that the amplitude of the oscillation is only known up to a normalizing 
 constant. In this particular plot (and in similar plots for the other cases) 
 we chose that constant in such a way as to make the oscillation visible to the reader.
 The right hand-side panels show the amplitude (top panel) and phase (bottom panel) of the radial velocity 
 as a function of the height in the atmosphere.

 As we can see from the top-right panel, the total amplitude (i.e., the amplitude derived 
 from fitting the radial velocity - black line)
 follows the  behavior of the parallel amplitude, derived from the fitting of 
 $V_\parallel$ and related to the contribution of acoustic waves (red line). 
It is, however, always smaller than the parallel amplitude because of the contribution from the magnetic waves, whose amplitude is derived from the 
 fitting of $V_\perp$ (gray line).
 In the bottom-right panel,
we see that the total phase (black line) follows relatively closely the parallel (acoustic) phase (red line) in the outermost layers, 
but diverges from it in the lower atmospheric region due to the increasing impact of the perpendicular (magnetic) phase. 
Despite this, the phase variation across the whole atmosphere is small.
The left panel of Fig. \ref{fig:atm_Am_ph_1} gives us also an idea of the behavior of the phase.
In the present case we can confirm the small variation of the phase with height, seen in the slight shift of the zeros to the right, as we move towards higher atmospheric layers.
 
This is a clear case in which the phase variation in the radial velocity results from the competition between the  acoustic and magnetic components that enter the integral, rather than from an actual phase variation in either of them.
 The perpendicular phase is always constant, 
 due to the standing nature of the magnetic waves, while the parallel phase, 
 related with the acoustic component, is constant due to the pole-on view, 
that favours the area 
 where the standing waves are located, and the low value of the frequency that guarantees that the standing acoustic waves occupy a larger 
 area in the visible disk of the star than the acoustic running waves.

 \subsection{Case 2}
 
\begin{figure}
	\includegraphics[width=\columnwidth]{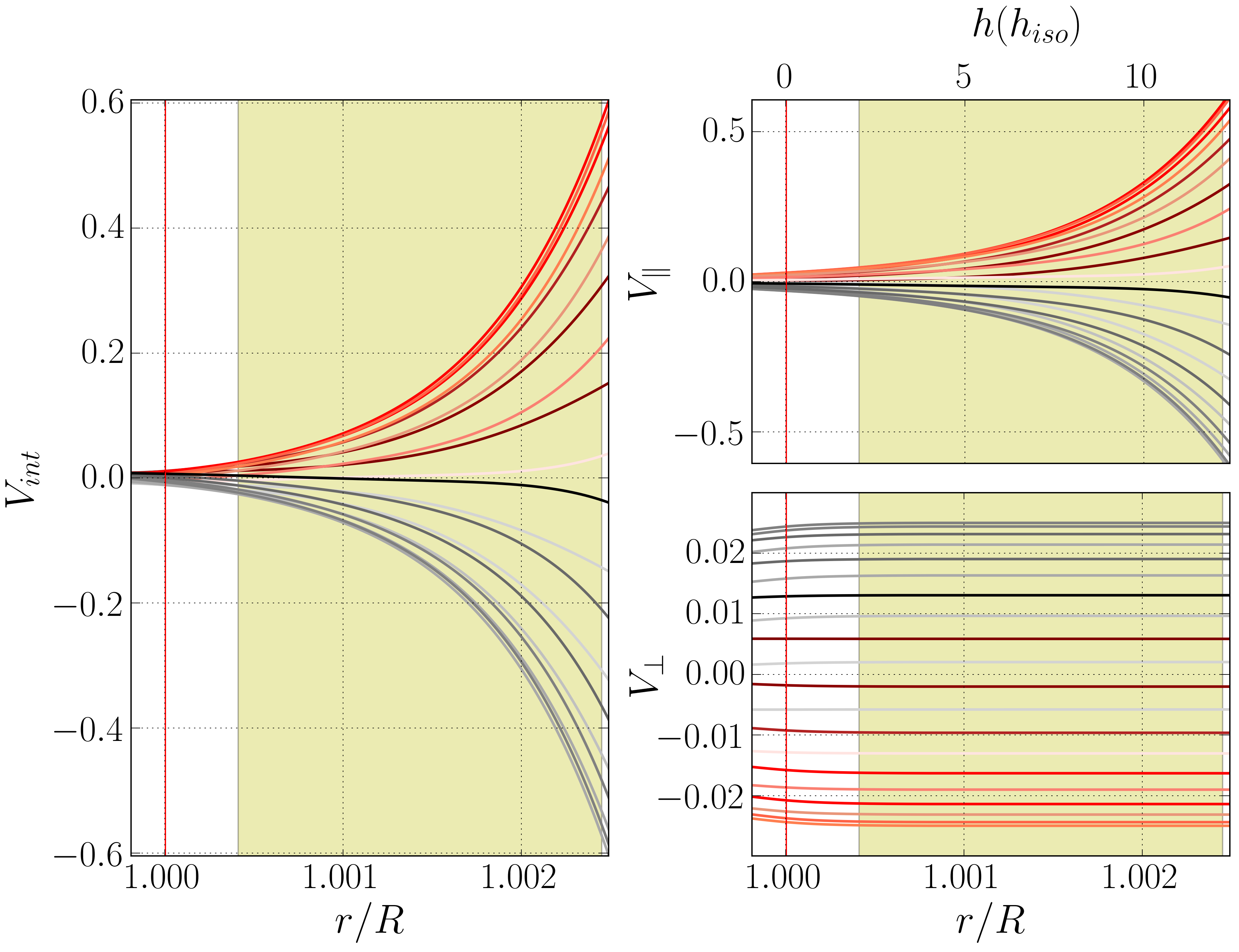}
    \caption{The same as Fig. \ref{fig:intRV_1} but for a mode of frequency $2.2$~mHz and degree $l=1$, 
    a magnetic field of $2$~kG, and an observer pole-on. 
    }
    \label{fig:intRV_2}
\end{figure}

 For the second case we consider a frequency bellow the acoustic cut-of, but close to it.
 We have chosen a mode with a frequency of $2.2$ mHz, degree $l=1$, and an observer pole-on.
 The angle at which the critical frequency becomes smaller than $2.2$ mHz is $\alpha_{cr}=33^{\circ}$ (cf. Table~\ref{tab:cases}).  
 The radial velocity is shown on the left panel of Fig. \ref{fig:intRV_2},
 and the contributions to it from the components of the velocity parallel and perpendicular to the magnetic field
 are shown on the right panels, in the same way as for the previous case. 
 We see, from the right panels, the exponential behavior of the acoustic wave's contribution (top) in the atmosphere of the star, 
 and the constant amplitude of the magnetic wave's contribution (bottom),
 but this time the amplitudes of the two contributions 
 differ more significantly in the high atmosphere.   
 This is because in the present case the fraction of the visible disk covered with  acoustic running waves is larger than in case 1. 
 Since the amplitude of the displacement, hence also of the velocity, 
 increases faster with height for running acoustic waves 
 than for standing acoustic waves (as discussed in section \ref{subsec:solut}), 
in the present case the acoustic contribution to the integral of the radial velocity becomes more dominant in the high atmosphere.
 
 \begin{figure}
	\includegraphics[width=4.01cm]{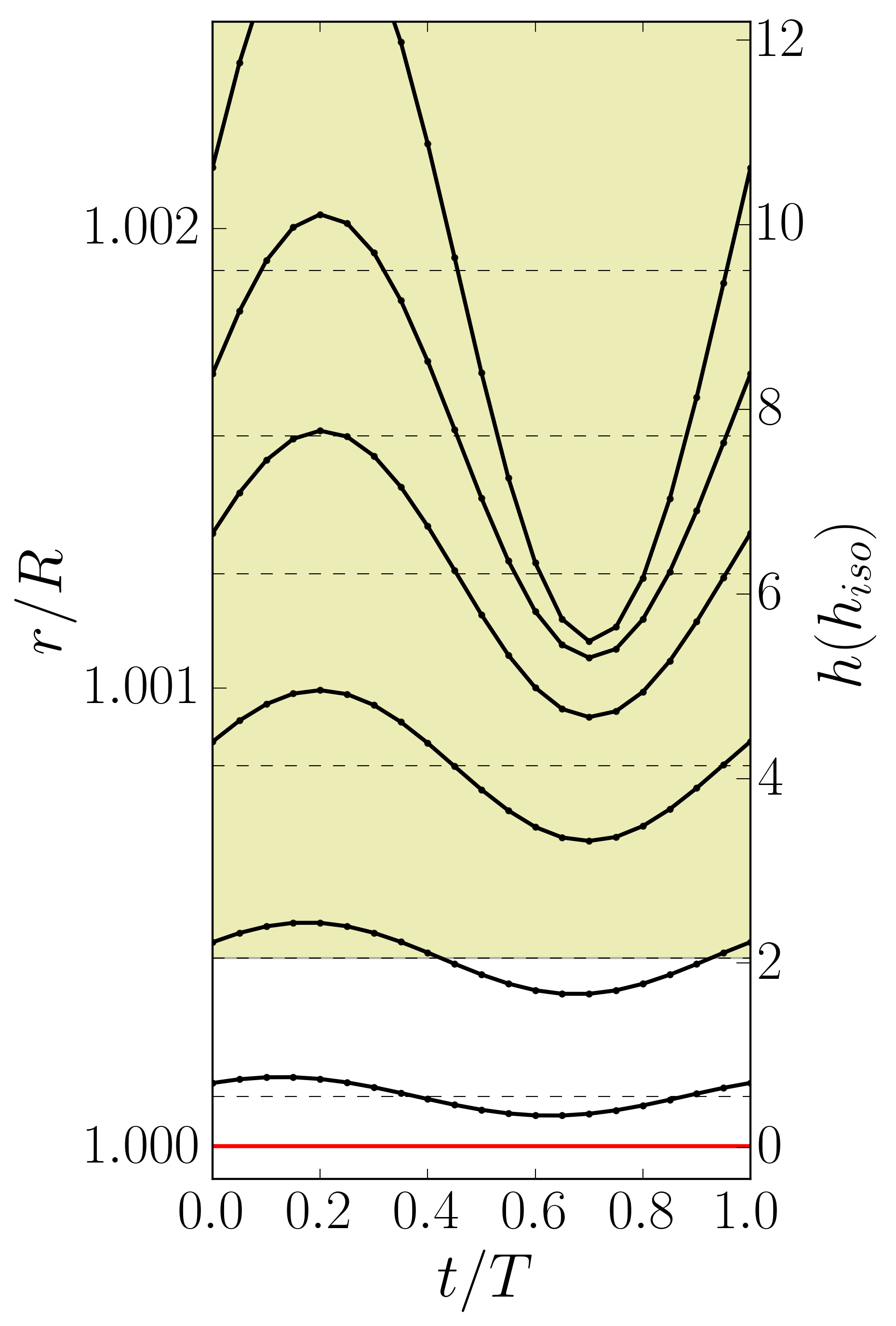}
    \includegraphics[width=4.2cm]{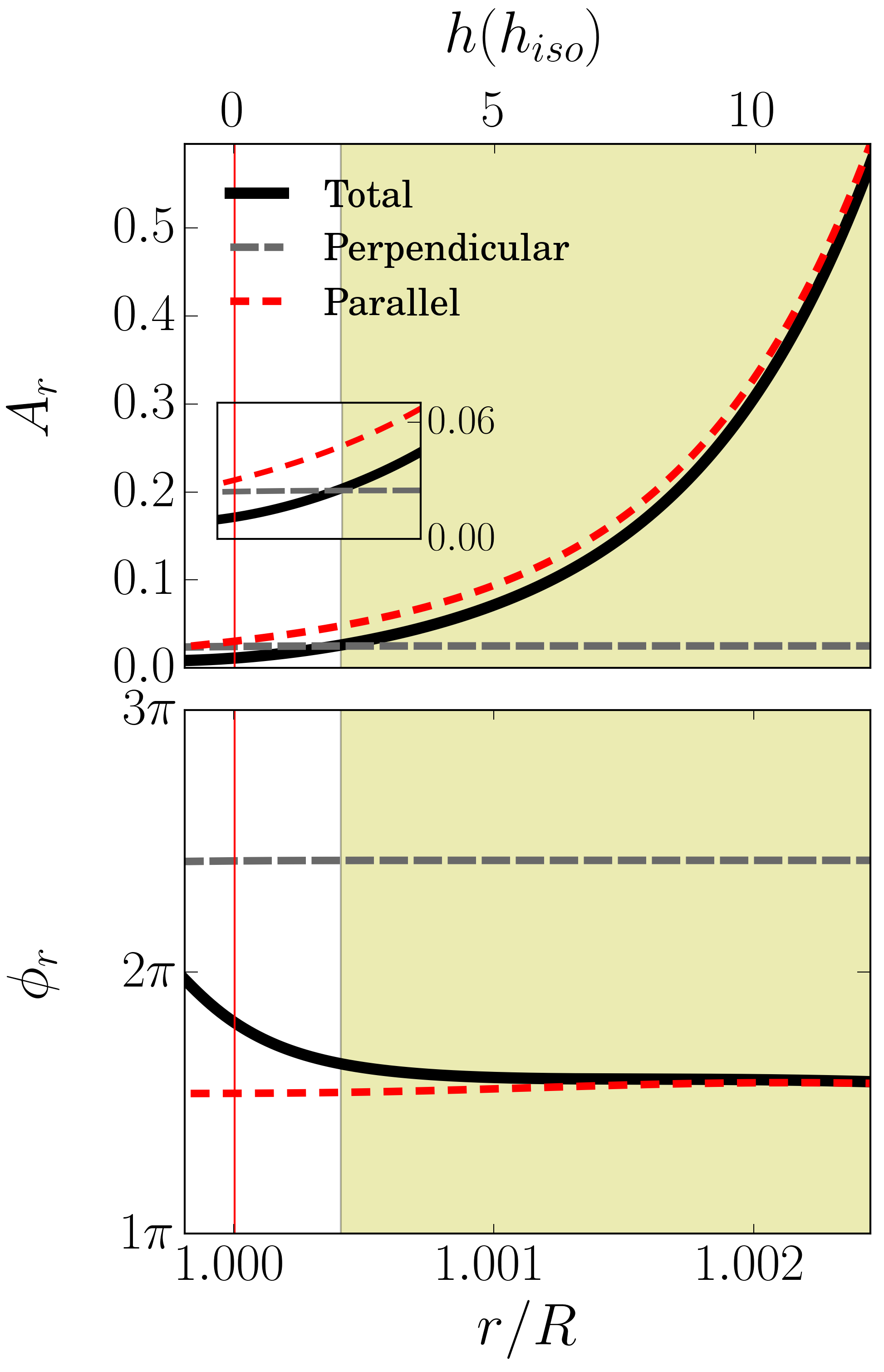}
	\caption{The same as Fig. \ref{fig:atm_Am_ph_1} but for a mode of frequency  $2.2$~mHz and degree $l=1$, 
	 a magnetic field of $2$~kG,  and an observer pole-on.  The
         close-up shows the behaviour of the amplitudes near the photosphere.
    }
    \label{fig:atm_Am_ph_2}
\end{figure}
   
 The amplitude and phase of the radial velocity are shown in Fig. \ref{fig:atm_Am_ph_2}, 
 in the same manner as in the previous case. 
In the outermost layers the total amplitude (black line, top-right panel) follows the acoustic wave's contribution (red line), 
 but, when moving towards lower atmospheric layers the magnetic wave's contribution (gray line) becomes increasingly important.
 This behavior can be seen also in the phase (bottom-right panel),
 as the total phase (black line) follows the phase from the acoustic wave's contribution 
 (red line) in the outer layers, but approaches the phase of the magnetic wave's (gray line) contribution deeper in.
 Unlike in the previous case, here we can see a very small variation in the phase of the acoustic 
 contribution (red line) which is due to the higher frequency of the mode considered that results 
 in a more significant contribution of acoustic running waves
to the integral of the parallel component. 
Nevertheless, the dominant phase variation in the radial velocity (black line) 
results from the competition between the contributions to the radial velocity integral of the parallel (acoustic) and perpendicular (magnetic) velocity components,
as in the previous case.
 
Due to the small variations in the phase, a small shift in the zeros can also bee seen in Fig. \ref{fig:atm_Am_ph_2}, left panel, 
when looking at different atmospheric heights.

\subsection{Case 3}

\begin{figure}
	\includegraphics[width=\columnwidth]{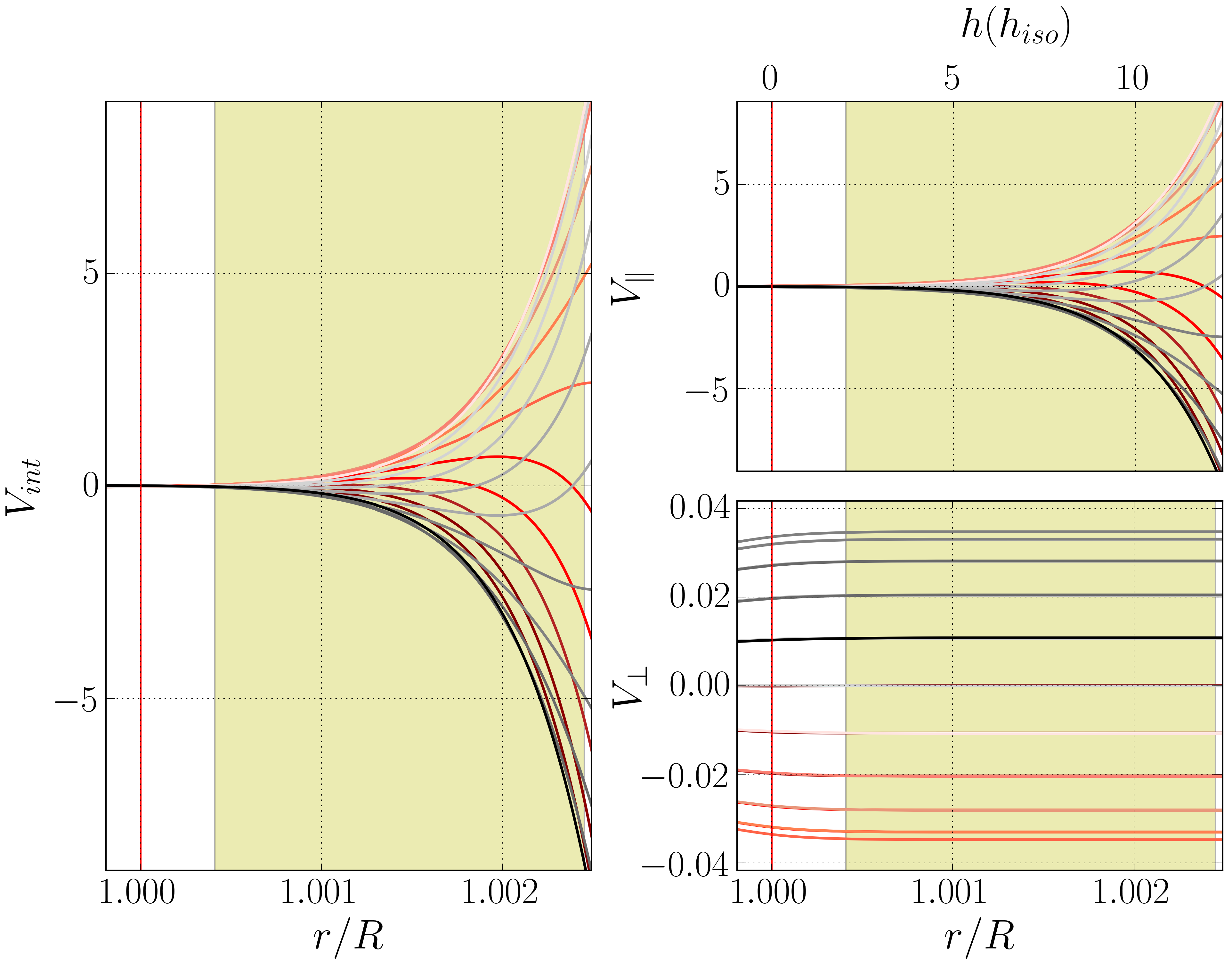}
    \caption{The same as Fig. \ref{fig:intRV_1} but for a mode of frequency $2.7$~mHz and degree $l=1$,
    a magnetic field of $2$~kG,
   and an observer pole-on. 
    }
    \label{fig:intRV_3}
\end{figure}

  The third case is one in which the phase is found to be more significantly variable.
  It is the last one we present with a pole-on observer and it concerns a mode with a frequency of 2.7 mHz, 
  which is above the acoustic cut-off (cut-off frequency of the star in Table \ref{tab:star}), 
  and a degree of $l=1$ (Table \ref{tab:cases}).
  The radial velocity for this case is shown in Fig. \ref{fig:intRV_3}, left panel.
  Since the mode frequency is above the acoustic cut-off, the acoustic running waves are present in the full visible disk.
  Due to the faster increase with height of acoustic running waves,
  the amplitudes of the acoustic wave's contribution (top-right panel) 
  and magnetic wave's contribution (bottom-right panel) 
 differ by 2 orders 
 of magnitude in the outermost layers, leading to a total dominance of the acoustic waves in that part of the atmosphere.
  
\begin{figure}
	\includegraphics[width=4.01cm]{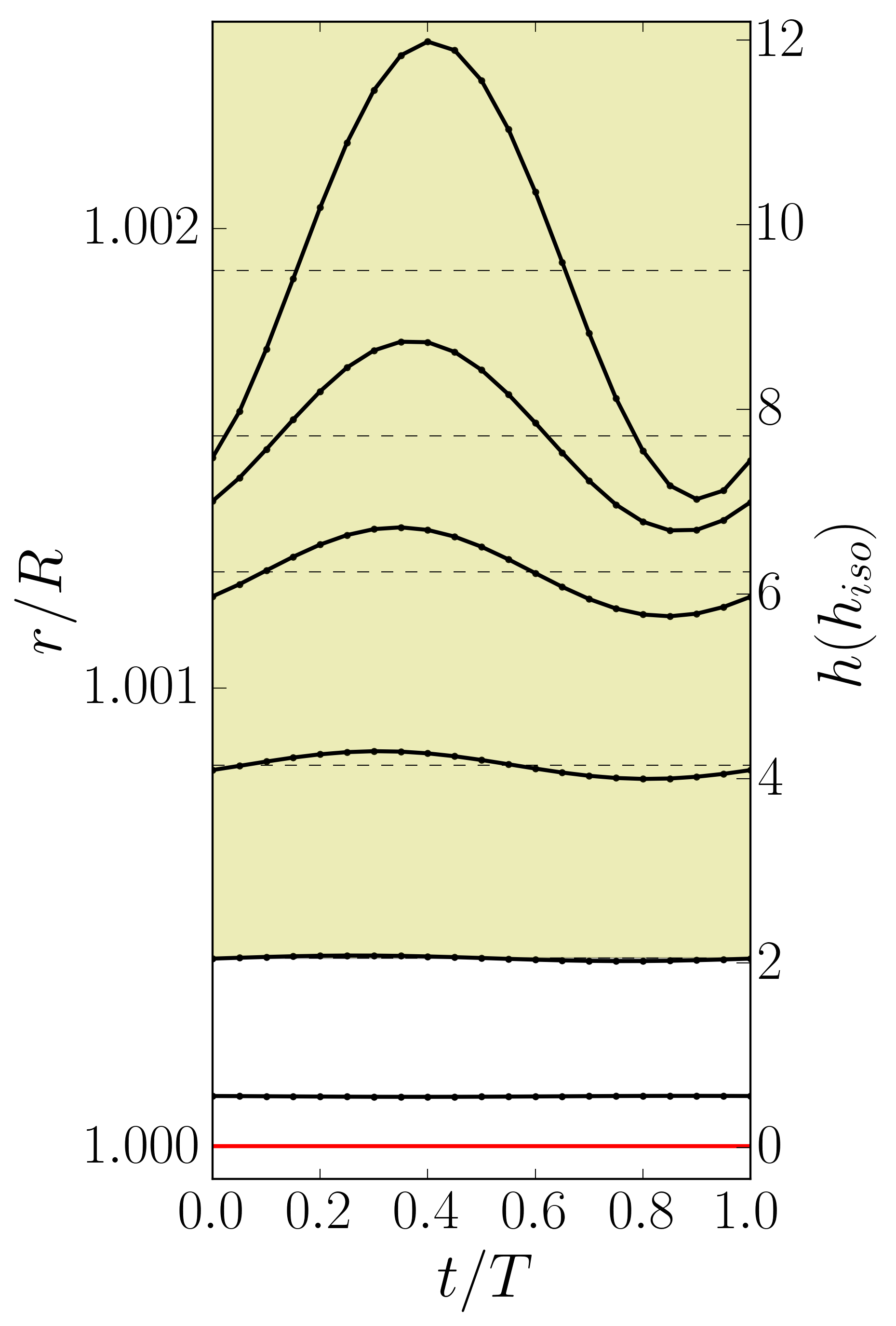}
    \includegraphics[width=4.2cm]{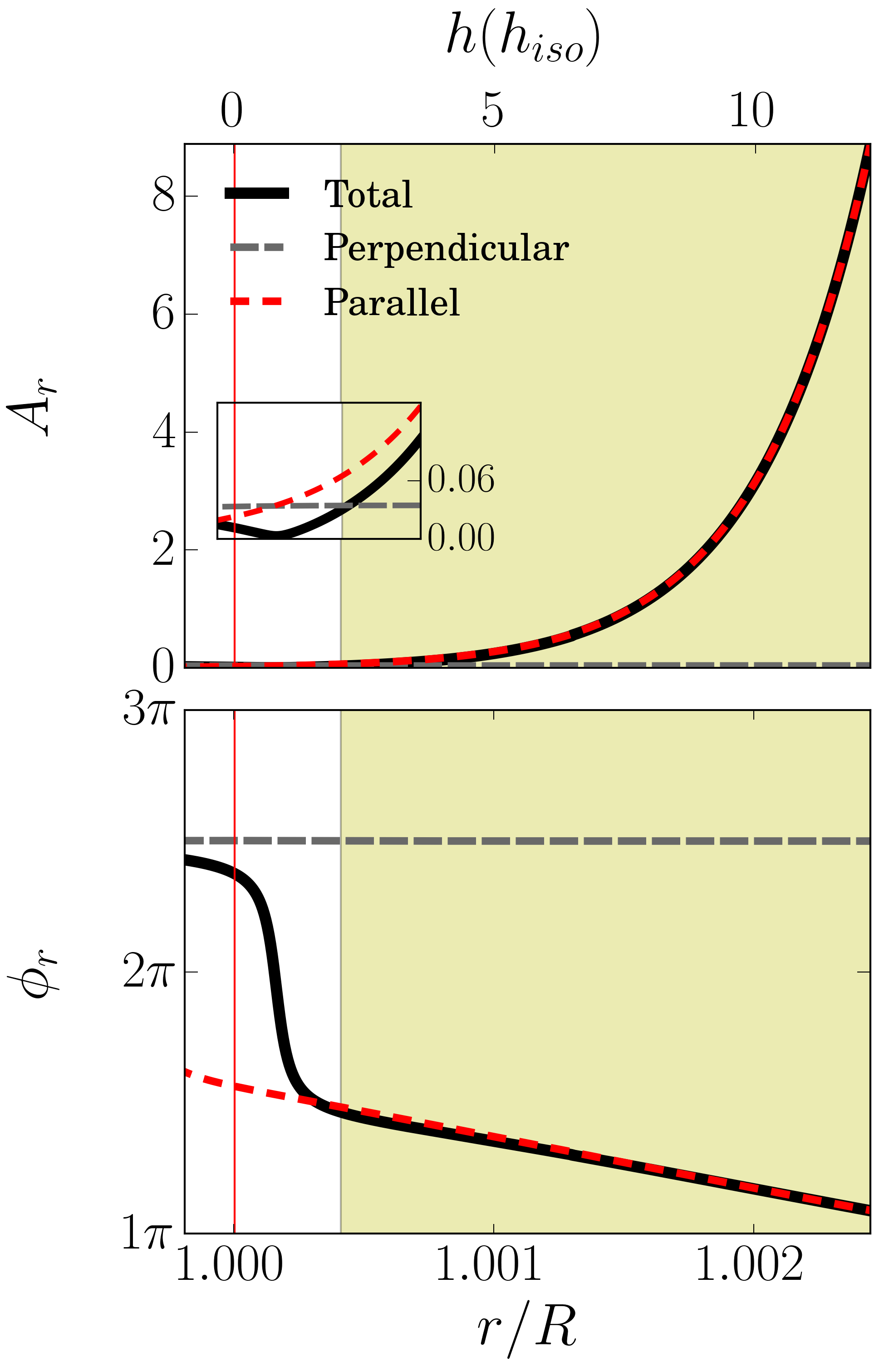}
	\caption{The same as figure \ref{fig:atm_Am_ph_1} but for a mode of frequency $2.7$~mHz and degree $l=1$, 
	  a magnetic field of $2$~kG,
      and an observer pole-on.   The close-up shows the behaviour of the amplitudes near the photosphere. }
    \label{fig:atm_3}
\end{figure}

  The variations in the amplitude and phase for this case are shown in Fig. \ref{fig:atm_3}, 
  top- and bottom-right panels, respectively.
 The variations in the total amplitude and the total phase (black lines) show the 
  dominance of the acoustic wave's contribution (red line) throughout the isothermal atmosphere. 
  In that region we can see a significant variation of the parallel phase caused by the running 
  acoustic waves.
  In the inner atmosphere, where the magnetic and acoustic contributions have the same order 
  of magnitude, we can identify a crossing between the acoustic (red line) and the magnetic 
  wave's contributions (gray line). 
 Together with the abrupt variation in the total phase (black line) this marks the transition between 
 the dominance of the two types of waves in the radial velocity integral. 
 Because they are out of phase, this crossing generates an apparent node in the inner atmosphere.

  Looking at Fig. \ref{fig:atm_3}, left panel, the variation in the phase can be clearly seen,
  as a shift in the zeros of the oscillations when comparing different atmospheric heights.

\subsection{Case 4}
 
\begin{figure}
    \includegraphics[width=\columnwidth]{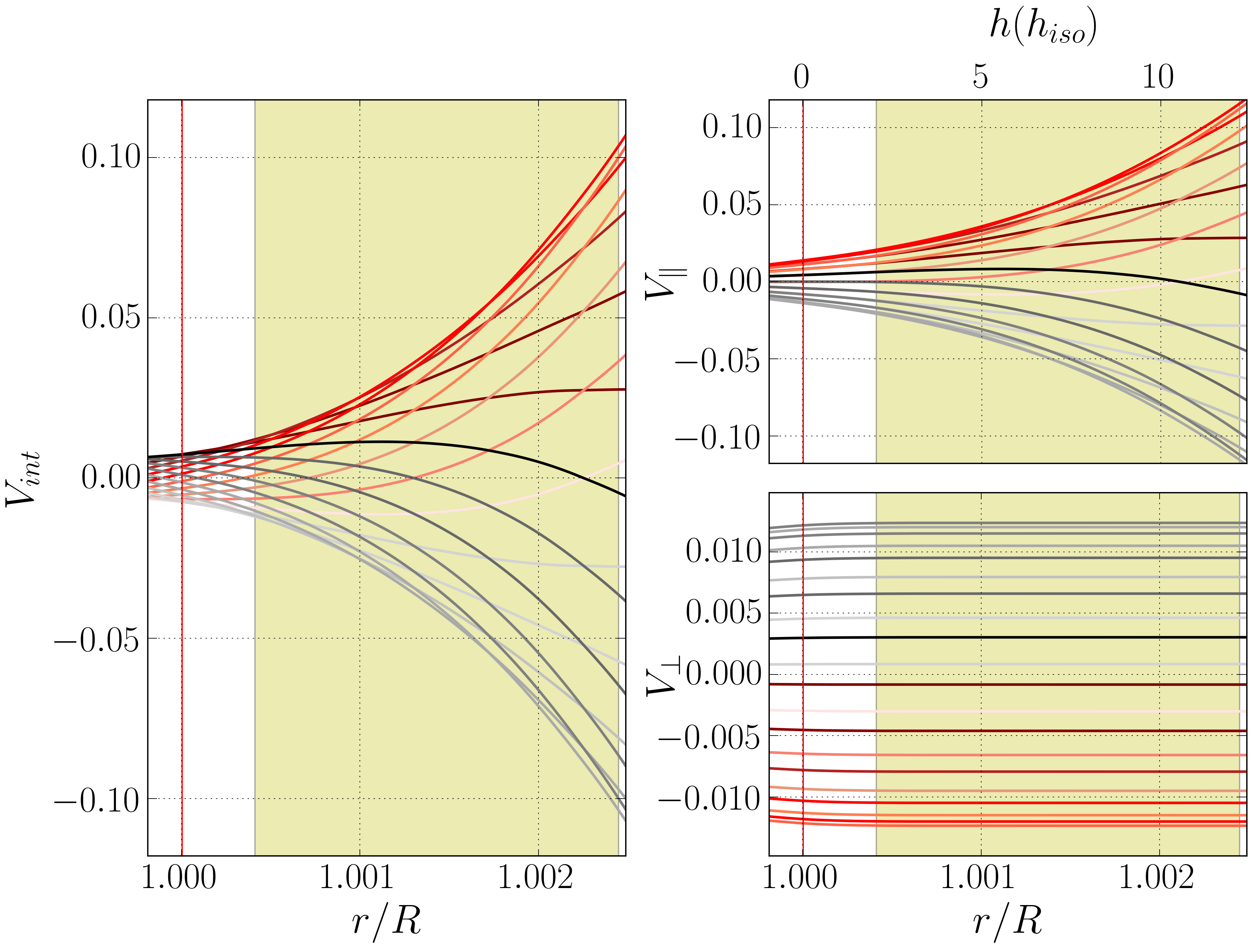}
    \caption{The same as Fig. \ref{fig:intRV_1} but for a mode of frequency $1.7$~mHz and degree $l=0$, 
    a magnetic field of $2$~kG,
    and an observer equator-on. 
    }
    \label{fig:intRV_4}
\end{figure}

\begin{figure}

	\includegraphics[width=4.01cm]{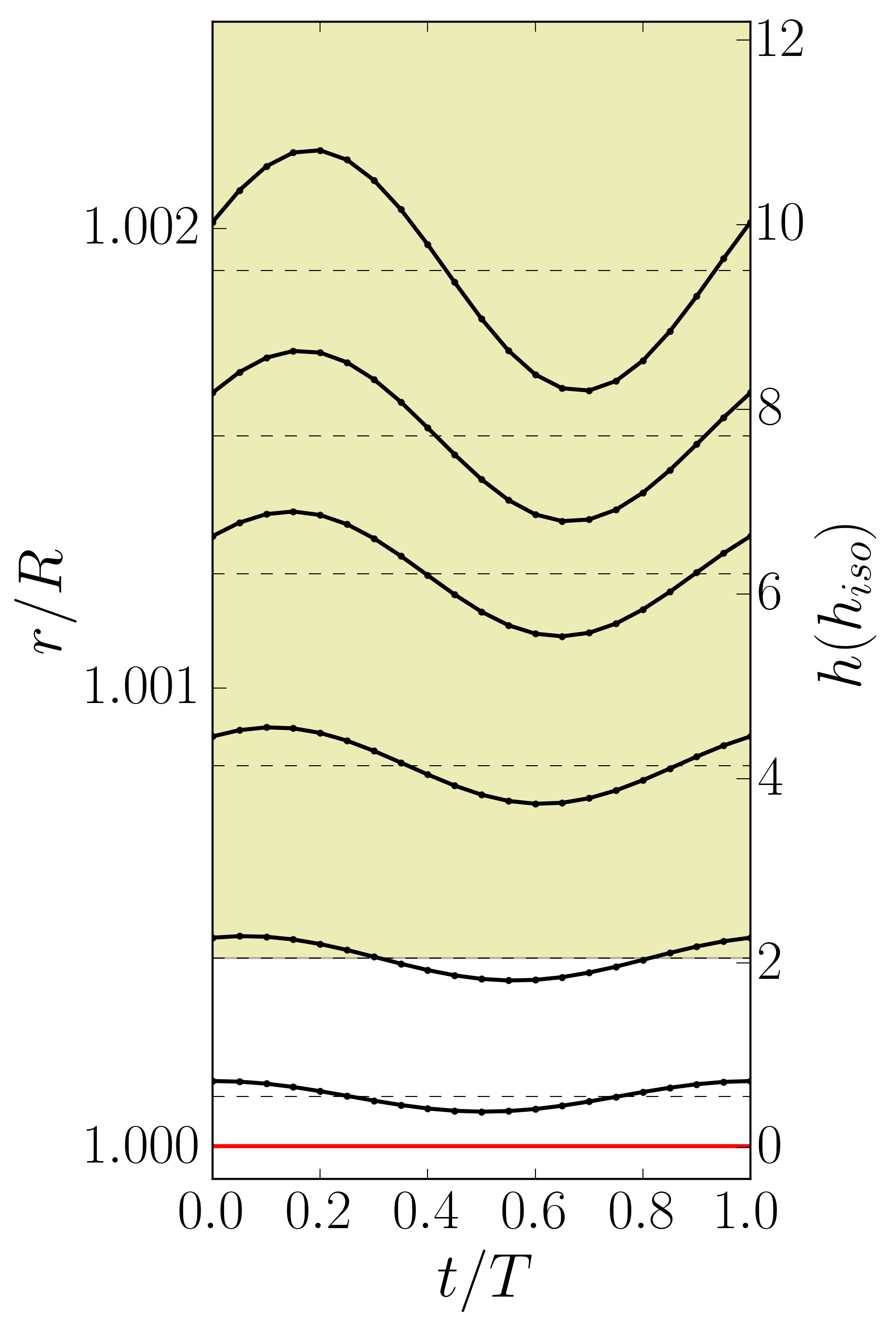}
    \includegraphics[width=4.2cm]{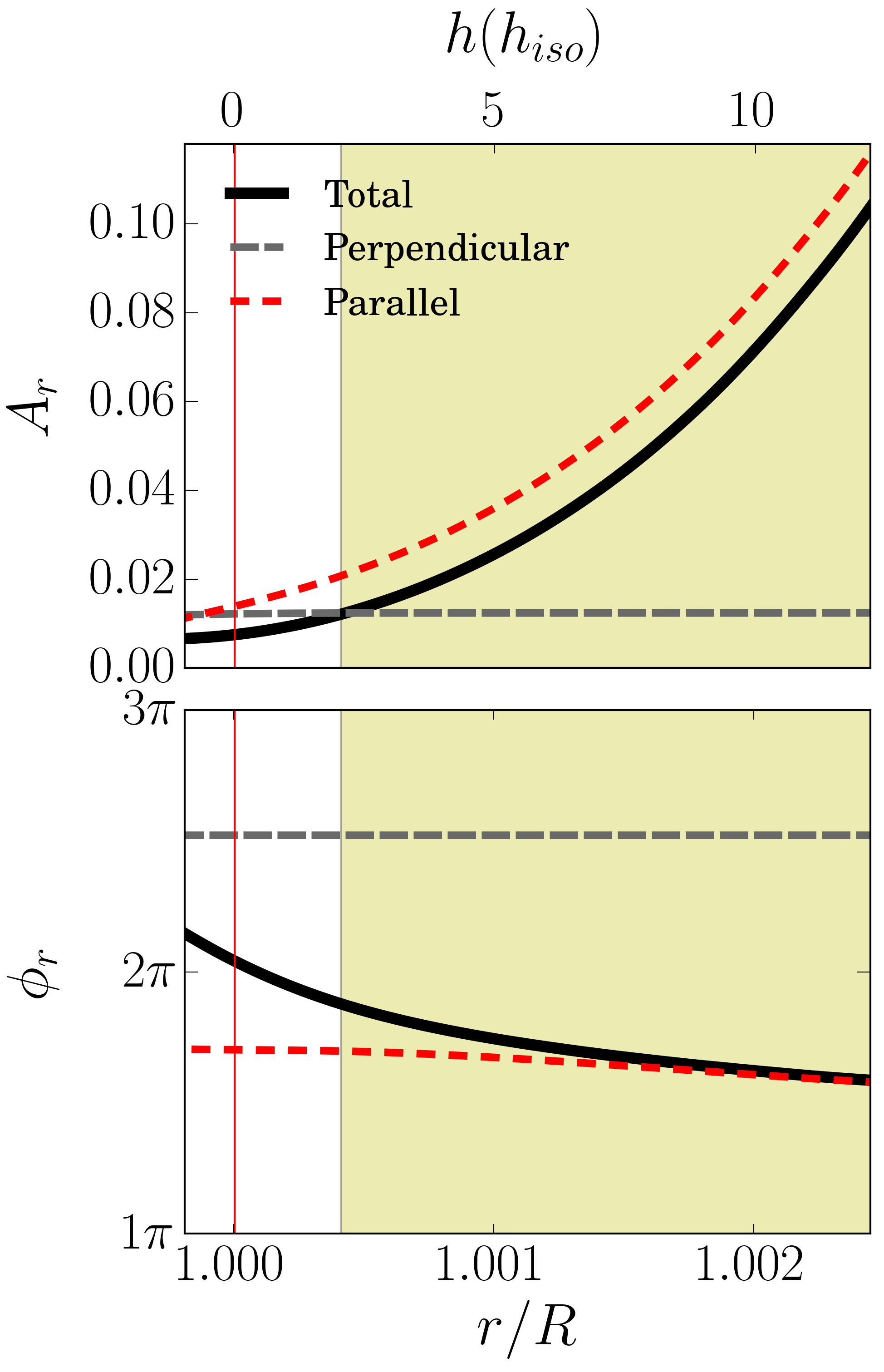}
	\caption{The same as Fig. \ref{fig:atm_Am_ph_1} but for a mode of frequency  $1.7$~mHz and 
	  degree $l=0$, a magnetic field of $2$~kG, and an observer equator-on. 
    }
    \label{fig:atm_4}
\end{figure}

 The case 4 corresponds to an observer located equator-on and a mode with a frequency of $1.7$~mHz and degree $l=0$ (Table \ref{tab:cases}).
 The radial velocity for this case is shown in Fig. \ref{fig:intRV_4}, left panel.
The amplitudes of the acoustic and magnetic waves' contributions (Fig. \ref{fig:intRV_4}, right panels)
are of the same order of magnitude in the lower part of the atmosphere, just as was found for case 1, although here the two contributions become comparable near the photosphere, as seen from Figure 13, right upper panel. But the main difference with respect to case 1 is the behavior of the phase.
  Looking at Fig. \ref{fig:atm_4}, lower right panel,
   we see that similarly to case 1 the total phase (black line) follows the parallel phase (red line) in the outermost layers, and diverges from it as one approaches deeper regions of the atmosphere, due to the influence of the magnetic waves.
However, contrary to case 1, the parallel phase (red line)
now varies with depth,
due to the contribution of the acoustic running waves that are concentrated towards the equator, 
where the observer is positioned. As a result, 
the total phase varies also in the outer atmospheric region. 
The phase variation can be seen also in the left panel of Fig. \ref{fig:atm_4}.

 \subsection{Case 5}
 
 The second case with an observer equator-on is for a mode with a frequency of $2.2$~mHz and
 degree $l=0$ (cf. Table \ref{tab:cases}).
 
\begin{figure}
    \includegraphics[width=\columnwidth]{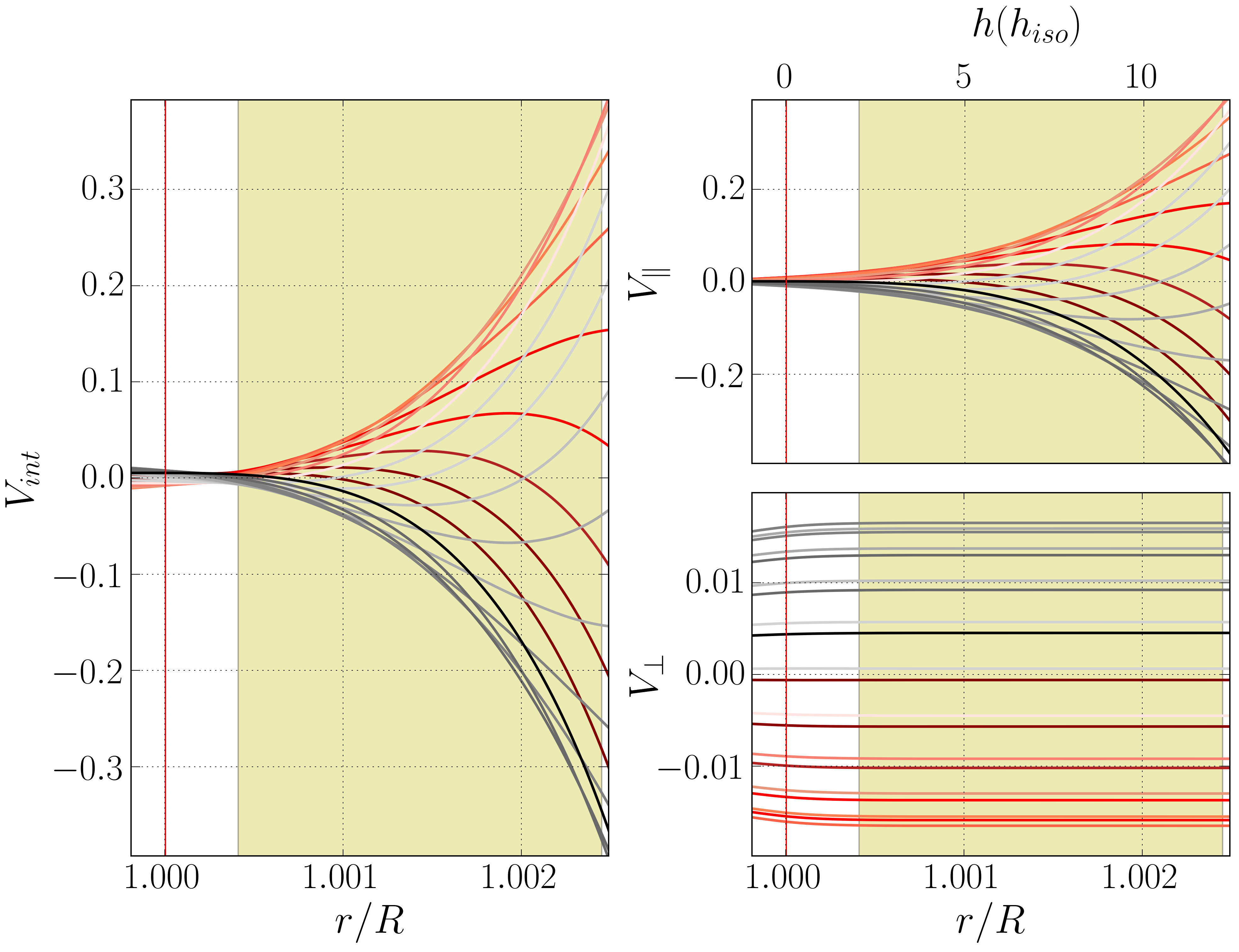}
    \caption{The same as Fig. \ref{fig:intRV_1} but for a mode of frequency  $2.2$~mHz and degree $l=0$, 
    a magnetic field of $2$~kG,
     and an observer equator-on.}
    \label{fig:intRV_5}
\end{figure}

 The radial velocity is shown in Fig. \ref{fig:intRV_5}, left panel,
 and the acoustic and magnetic waves' contributions are shown in  Fig. \ref{fig:intRV_5}, top- and bottom-right panels,
 respectively. In the outer atmospheric layers, the two contributions differ by 1 order of magnitude,
 just as in case 2, with the same frequency but a pole-on observer.
 Thus, the acoustic waves are dominant in those layers. 
 While this is similar to case 2, here we can see a modulation with height of the exponential behavior in the atmosphere. 
 This is due to the fact that in this case the observer is looking more directly at the  acoustic running waves.

\begin{figure}
	\includegraphics[width=4.01cm]{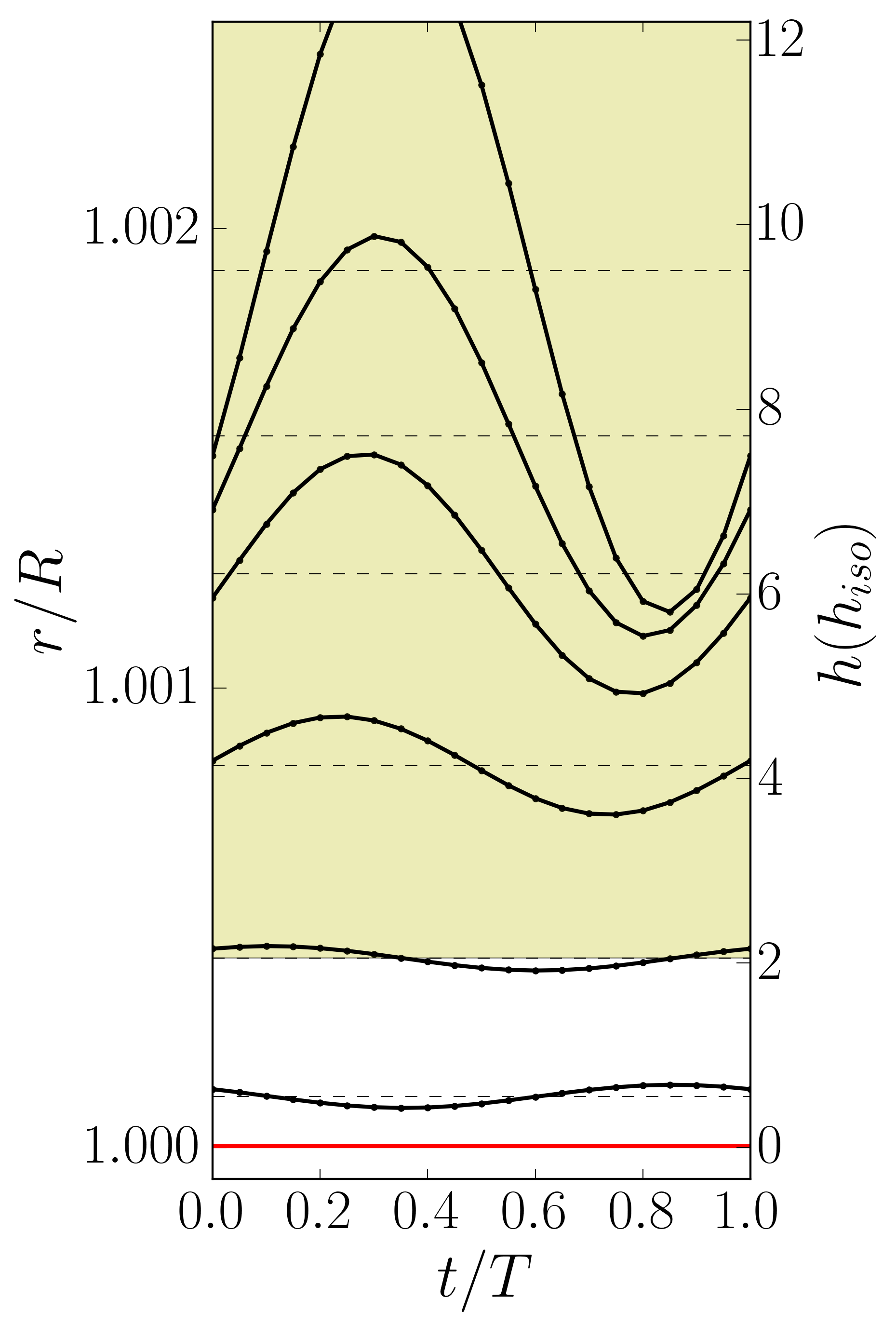}
    \includegraphics[width=4.2cm]{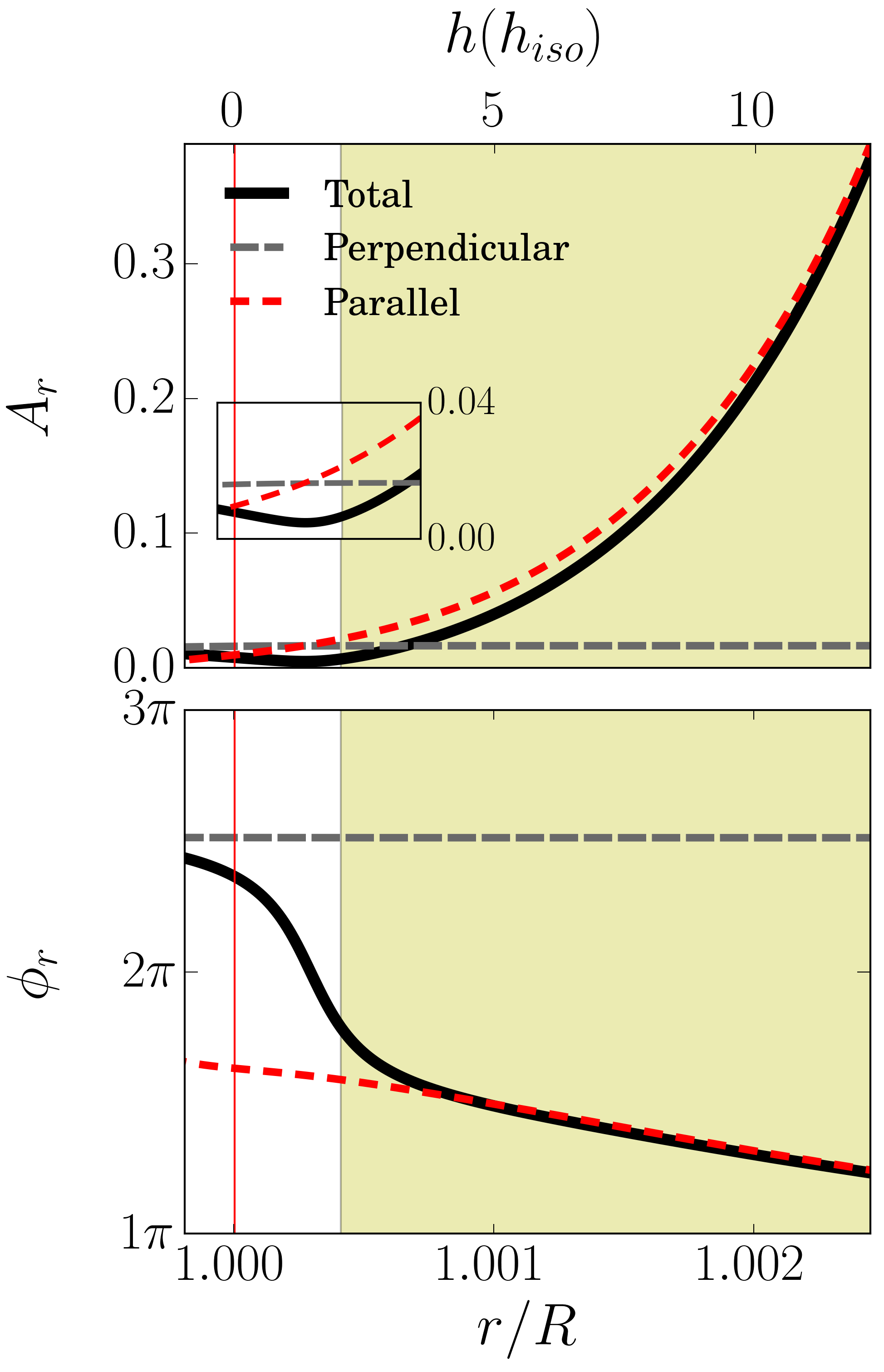}
	\caption{The same as Fig. \ref{fig:atm_Am_ph_1} but for a mode of frequency $2.2$~mHz and degree $l=0$, 
	a magnetic field of $2$~kG,
    and an observer equator-on. The
         close-up shows the behaviour of the amplitudes near the photosphere.
    }
    \label{fig:atm_5}
\end{figure}

 The amplitude and phase variations of the radial velocity are shown in Fig. \ref{fig:atm_5}, right panels, 
where again we have the total amplitude (black line, top panel) dominated by 
 the amplitude derived from the acoustic wave's contribution (red line, same panel) in the high atmosphere, 
and a total phase (black line, bottom panel) that changes from following the phase derived from the acoustic wave's  
 contribution in the high atmosphere (red line, same panel) to following the phase derived from the magnetic wave's contribution
   (gray line, same panel) in the inner layers of the atmosphere.
 
 Considering Fig. \ref{fig:atm_5}, left panel, 
 the change in the phase is evident and much greater than the phase variation seen in the case 2.
 This is again because with the pole-on view the observer is looking directly at the acoustic standing waves at the pole, 
 but with the equator-on view the observer is looking directly at the acoustic running waves in the equator.

\subsection{Case 6}

\begin{figure}
	\includegraphics[width=\columnwidth]{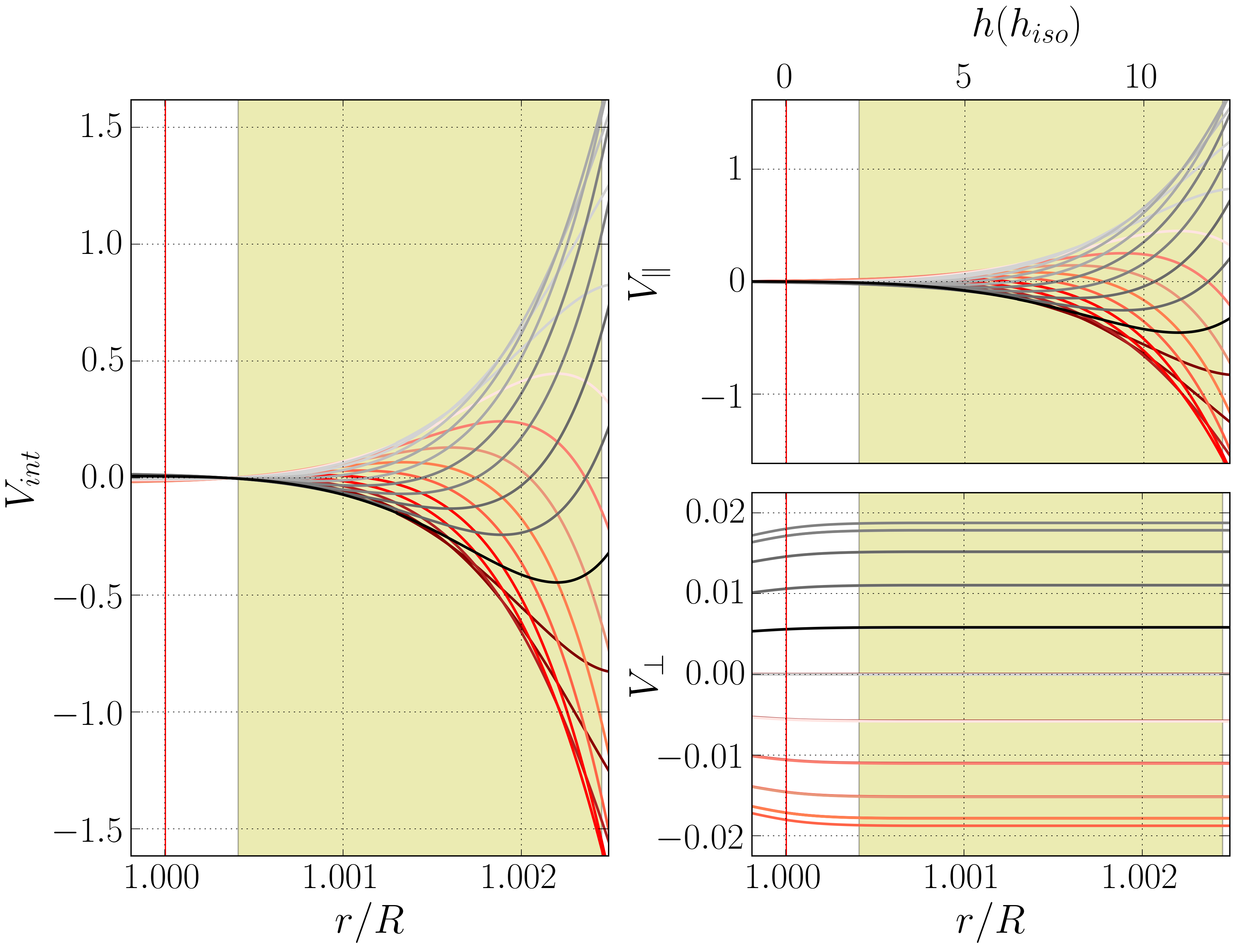}
    \caption{The same as Fig. \ref{fig:intRV_1} but for a mode of frequency $2.7$~mHz and degree $l=0$, 
    a magnetic field of $2$~kG,
     and an observer equator-on. 
    }
    \label{fig:intRV_6}
\end{figure}

 This case is for a mode with a frequency of $2.7$ mHz, above the acoustic cut-off, and a degree $l=0$, and an equator-on observer
(cf. Table \ref{tab:cases}).
 The radial velocity, seen in Fig. \ref{fig:intRV_6}, left panel, 
 shows a fast exponential growth, which is modulated with height,
%
 since the acoustic running waves are present in the full visible disk.
 From inspection of Fig. \ref{fig:intRV_6}, right panels, we can see, as in the previous cases, 
 that the contribution from the acoustic wave to the radial velocity dominates throughout most of the atmosphere, 
 with the exception of the innermost layers.

\begin{figure}

	\includegraphics[width=4.01cm]{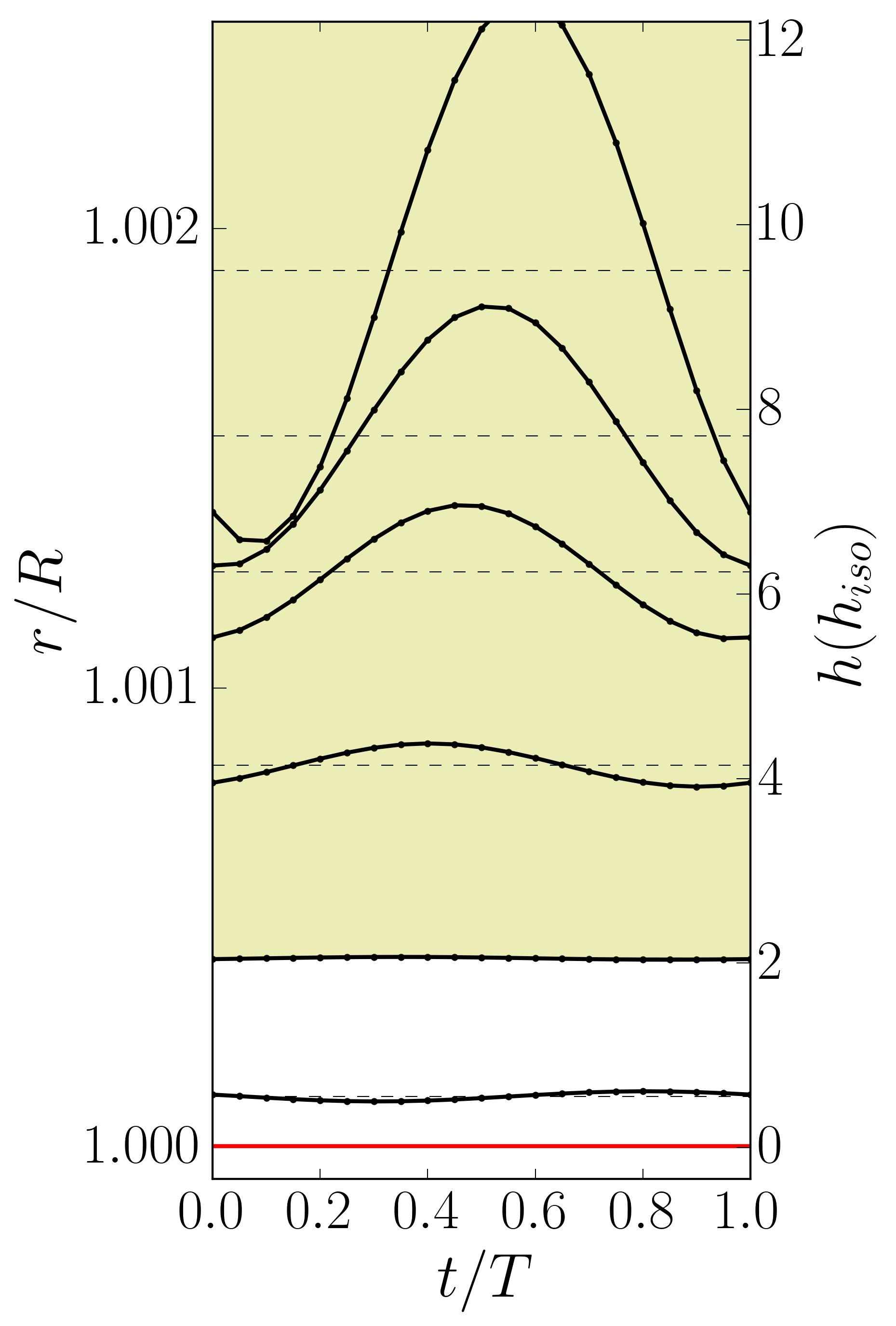}
    \includegraphics[width=4.2cm]{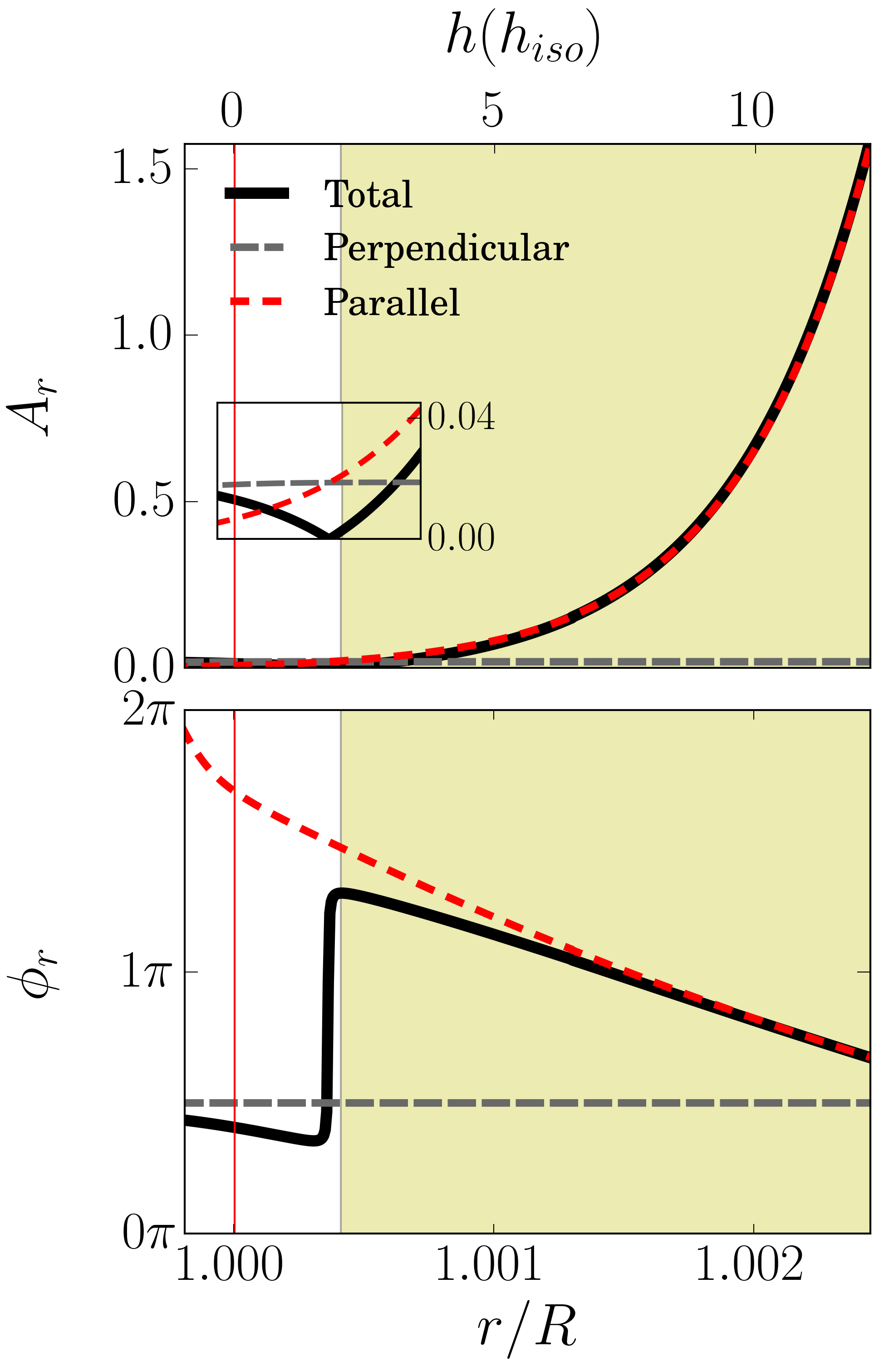}
	\caption{The same as Fig. \ref{fig:atm_Am_ph_1} but for a mode of frequency  $2.7$~mHz and degree $l=0$, 
	a magnetic field of $2$~kG,
     and an observer equator-on. The
         close-up shows the behaviour of the amplitudes near the photosphere.
    }
    \label{fig:atm_6}
\end{figure}

   Figure \ref{fig:atm_6}, right panels, shows the amplitude and phase variations for this case.
 We see a very significant phase variation (black line, bottom panel)
  in the isothermal atmosphere,
   which is mainly due to the contribution of the acoustic running waves (red line, same panel). 
   A significant phase variation was already seen in case 3, for the same frequency, with a pole-on observer, 
   but it is even more significant here. 
   This is because, 
   the acoustic running waves near the equator are propagating almost perpendicularly to the equator-on observer. 
   Therefore, their wavenumber projection into the line-of-sight direction is very large, 
   resulting in rapid height-variations in the center of the visible disk, 
   which contribute significantly to the radial velocity integral. 
   As before, in these layers
   the total amplitude (black line, top panel) is dominated by the acoustic wave's contribution (red line, same panel).
 In the inner atmosphere we can identify a jump of $\pi$, caused by the change in 
  the dominant contribution, from acoustic in the outer layers to magnetic in the inner layers.
A phase jump had already been seen in case 3, but in this case the change is sharper. 
This is because here the magnetic and acoustic contributions are completely out of phase. 
  
 Although we do not illustrate it here, we have verified that this particular apparent node for a mode of this frequency and degree l=0 can be seen from any observation angle, 
 although its exact position changes with the observer's view.
  
 The variation in the phase in the upper atmosphere is also very clear in Fig. 17 (left panel).

\subsection{Case 7}
 
The last case  that we will consider is selected to illustrate an apparent node in the middle of the isothermal atmosphere, 
 as first discussed by \citet{sousa2011understanding} 
 for a toy model of a full isothermal atmosphere.
 This case considers a mode with a frequency of $2.7$~mHz, 
 that is above the acoustic cut-off, and degree $l=0$, and a magnetic field of $B_p=2$~kG (cf. Table \ref{tab:cases}). 
 Moreover, the observer is pole-on, and we assume the elements are concentrated around the equator, 
 in the region defined by $53^o<\theta<127^o$, which are, thus, considered as limit angles, $\theta_i$ and $\theta_f$, 
 for the integration in eq. (\ref{eq:vh}).
 
\begin{figure}
	\includegraphics[width=\columnwidth]{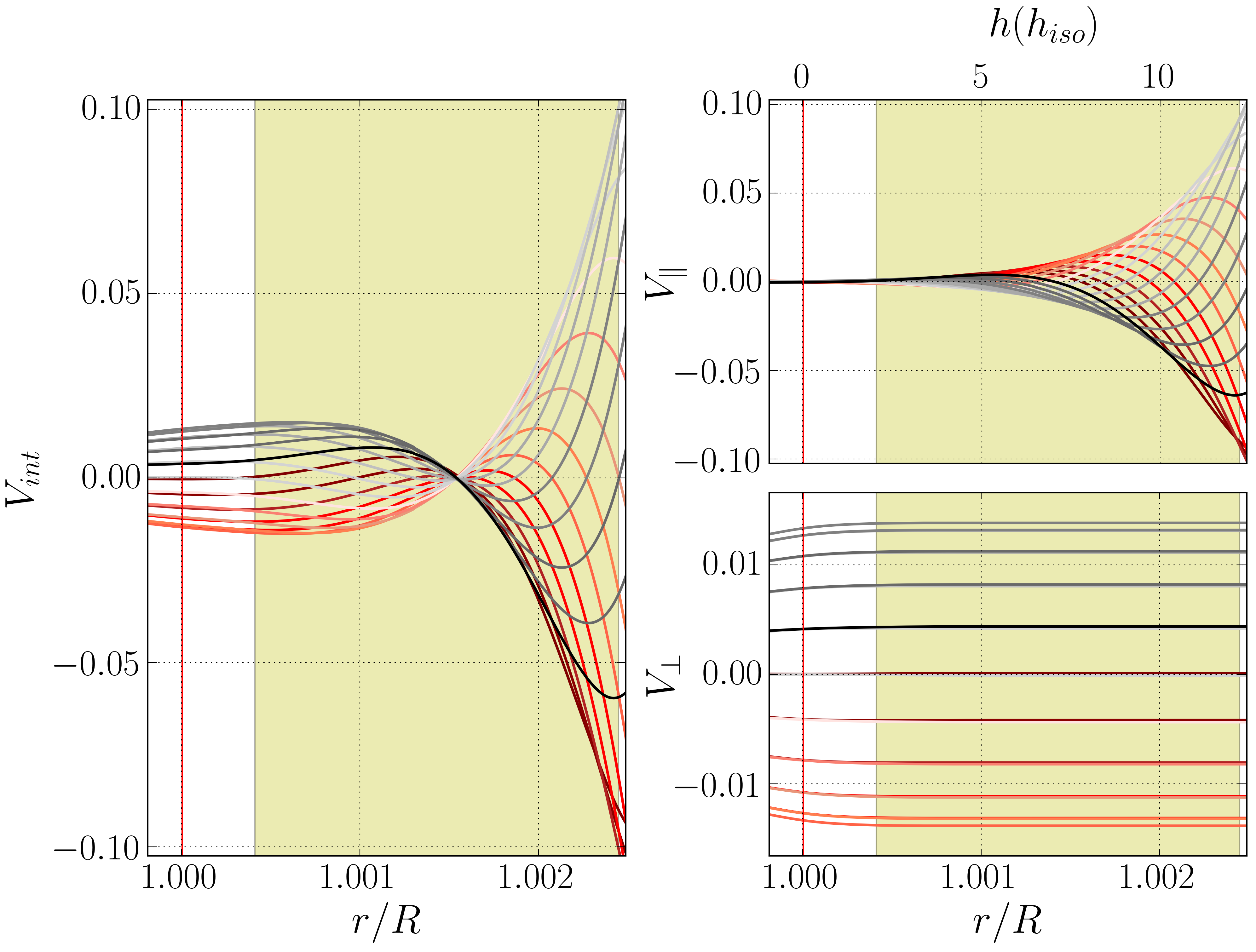}
    \caption{The same as Fig. \ref{fig:intRV_1} but for a mode of frequency $2.7$~mHz and degree $l=0$, 
    a magnetic field of $2$~kG,
     and an observer equator-on. The integration, for this case only, is in the region defined by $53^o<\theta<127^o$.
    }
    \label{fig:intRV_7}
\end{figure}

 The radial velocity for this case is shown in Fig. \ref{fig:intRV_7}, left panel.
 In the middle of the isothermal atmosphere we can see a sudden change in the radial velocity
 which looks somewhat similar to what one would expect in the presence of a node.  
 
 From the inspection of the right panels of the same figure, 
 we notice that in this case the acoustic and magnetic waves' contributions 
 are overall of  the same order of magnitude. Thus, we can recognize in the radial velocity shown in the left panel,  
 the exponential behavior of the acoustic waves in the upper atmosphere, 
 but also, the constant behavior of the magnetic waves in the inner atmosphere.
And like in case 6, we can see that the acoustic and magnetic waves' contributions are out of phase,
 as seen by the fact that for a given time (given color line in the right panels), 
 the acoustic (top panel) and magnetic (bottom panel) contributions have opposite sign. 
This leads again to a cancellation in the integral defining the radial velocity and is the cause of the apparent node.
  
 For the stellar model used in this paper, we find this type of apparent nodes in the higher atmospheric layers when considering elements distributed around the equator. They are seen from any position, and, 
 more commonly, for even-degree l modes.
 
\begin{figure}
	\includegraphics[width=4.01cm]{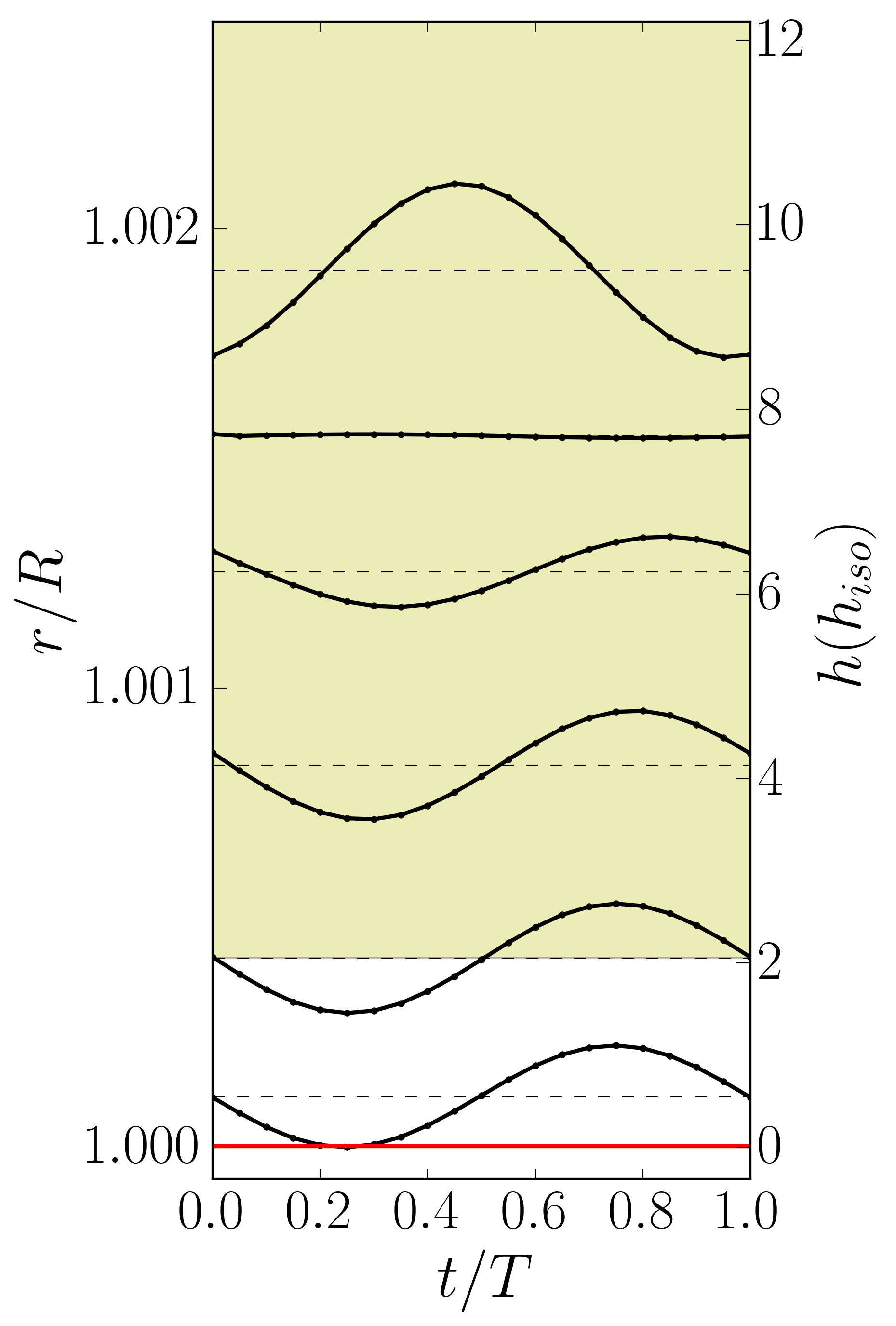}
    \includegraphics[width=4.2cm]{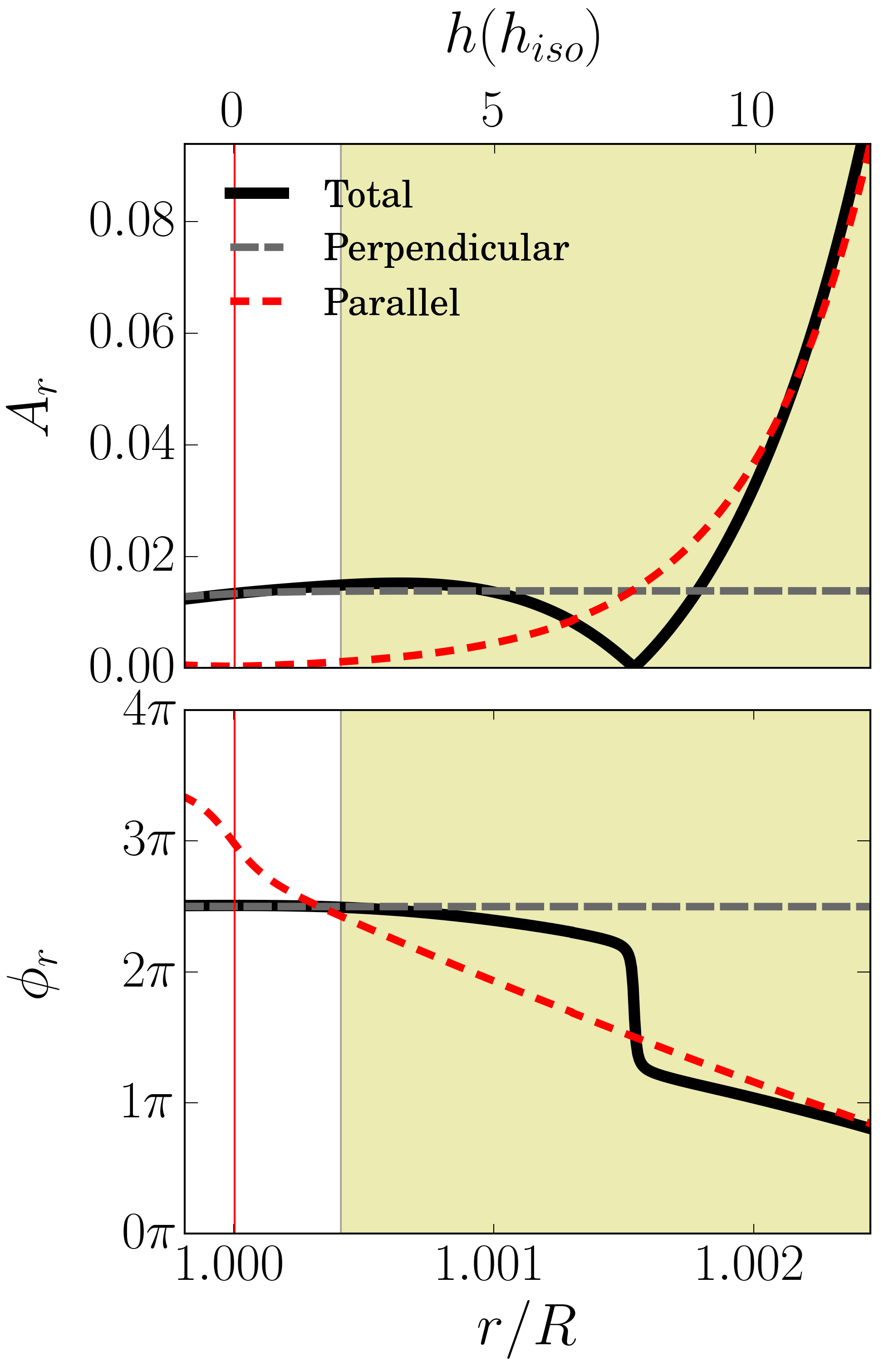}
	\caption{The same as Fig. \ref{fig:atm_Am_ph_1} but for a mode of frequency  $2.7$~mHz and degree $l=0$, 
	a magnetic field of $2$~kG,
     and an observer pole-on. The integration, for this case only, is in the region defined by $53^o<\theta<127^o$.}
    \label{fig:atm_7}
\end{figure}

 The amplitude and phase variations are shown in Fig. \ref{fig:atm_7}, right panels. 
 We see the total amplitude (black line, top panel) changing from behaving similarly to the amplitude of 
 the acoustic wave's contribution (red line, same panel) 
 in the upper atmosphere to behaving like the amplitude of the magnetic wave's contribution 
 (gray line, same panel)  in the inner atmosphere. Moreover,  
 because the magnetic and acoustic waves' contributions are similar in magnitude but with opposite sign, 
 at some point in the isothermal atmosphere the total amplitude decreases, going through a local minimum.
 At the same location we see the total phase varying by $\pi$ (black line, bottom panel). 
 These variations in amplitude and phase, as well as those found in cases 3 and 6 in the inner atmospheric layers, would, in an observational context, be interpreted as a presence of a node. 
 However, 
 this behavior is not caused by a node in a standing wave. 
 It's simply a visual cancellation effect between the acoustic and magnetic contributions to the radial velocity.

\section{Discussion and Conclusion}\label{sec:discu}

\subsection{General behavior of the amplitude and phase}\label{sec:discuwork}

Our results show that in general the amplitude increases rapidly with
height, due to the rapid increase of the amplitude of the acoustic
component. How significant the increase is depends also on the frequency.
As larger frequencies are considered, the radial velocity can reach greater amplitudes due to the increased presence of 
acoustic running waves in the atmosphere of the star. 
As for the phase behaviour, for frequencies below the
  acoustic cut-off the phase may vary due to a change in the type of waves that dominate
  the radial velocity integral. Moreover, depending on the position of
  the observer, the contribution of the
  acoustic waves in regions where the frequency is above the critical
  frequency, $\omega_c$, may
become dominant, resulting in a change of the phase due to the running
acoustic waves. When the frequency is above the acoustic cut-off, the
phase is found to vary regardless of the position of the observer.
Finally, we note that the position of the observer influences
  the phase behaviour not only because it determines the fraction of
  observed area where
  running waves are present, but also because the direction of the
  magnetic field around the equator makes the sound waves travel
  inclined to an observer that has an equator-on view,
making the projection of that component of the velocity field in the direction of the observer vary on short scales.

Concerning the contributions from the acoustic and magnetic components of the wave where these are decoupled, 
after inspecting the six cases with integration of the entire visible disk we can note that the acoustic waves dominate 
the behavior of the radial velocity in the upper atmosphere for  most of the cases.
This is explained by the difference in the amplitude  behavior of the acoustic and magnetic waves.
While the first has an exponential behavior in the atmosphere, the second has a constant behavior, making
the acoustic waves' contributions dominant in the outer layers of the atmosphere.

In the inner layers of the atmosphere we see a different
  scenario,  as
the magnetic waves start to have an influence, changing the amplitude and the total phase of the radial velocity.
This is the region where the change of dominance from acoustic to
magnetic waves' contribution occurs in our model, 
giving rise to a phase variation that in some cases may be abrupt enough to form an apparent node.
The position of this node, found when
integrating the whole visible disk, is expected to depend on  the place
where the magnetic and acoustic waves decouple (illustrated in
Fig. \ref{fig:pres} for the current model), since that decoupling
determines the relative amplitude of the two components, which beyond
that point have a different dependence on atmospheric height.  For
that reason, it is expected that the position of the node will be
different for models with different global properties (e..g, different
temperature).

Finally, we find that apparent nodes in the higher
  atmospheric layers appear often for spots or belts of elements in the equatorial area, 
when acoustic running waves are present. 
Exploring several frequencies and mode degrees we found that this phenomena can occur for any position of
the observer.
Also, for the node discussed in case 7 we have explored further configurations, 
changing the width of the belt and also the symmetry of the limits of integration,
and found that the apparent node remains present, 
showing only slight changes either in the minimum amplitude,
or in the atmospheric height position. However, it disappears when the interval over which the integration is performed is greater than $49^o<\theta<139^o$.

\subsection{Comparison with the observations}\label{sec:discuobs}
Our model shows that the radial velocity amplitude increases significantly 
(can reach one to two orders of magnitude) throughout the atmosphere. 
This increase is a direct consequence of the decrease in the density, 
as discussed in \citet{sousa2011understanding}, 
and it is most significant when the integral defining the radial velocity is dominated by the running acoustic waves. 
This is in agreement with the behaviour of the radial velocity amplitudes inferred from the observations, 
derived from absorption lines that are formed at different depths in the atmosphere \citep{ryabchikova2007pulsation,ryabchikova2007pulsation3D,kochukhov2008discovery,freyhammer20093d}.

In addition, our model shows that the phase variations throughout the atmosphere can take a variety of forms, 
that depend critically on the position of the observer and on the frequency of the modes. 
While in most cases the phase varies smoothly with height in the atmosphere,
in some cases the variations are sharper, taking place over relatively short distances. 
These sharper variations can be found both in the low and
  high  atmospheric regions, in our model, depending on the conditions. The latter case (e.g., Fig. \ref{fig:intRV_7} and \ref{fig:atm_7},
with sharp phase variations seen at densities between $\approx
10^{-9}$ and $10^{-11}$ gcm$^{-3}$), is of particular interest
when comparing with the observations.
Smooth, as well as sharp radial velocity phase variations are also commonly inferred from the spectroscopic time-series of roAp stars, 
particularly in the strongest pulsating lines that form high in the atmosphere, 
between optical depths of about log tau=-4 and -6 \citep{saio2010pulsations,saio2012pulsation}, corresponding to regions of low densities similar to that mentioned above.

 \begin{figure}

	\includegraphics[width=\columnwidth]{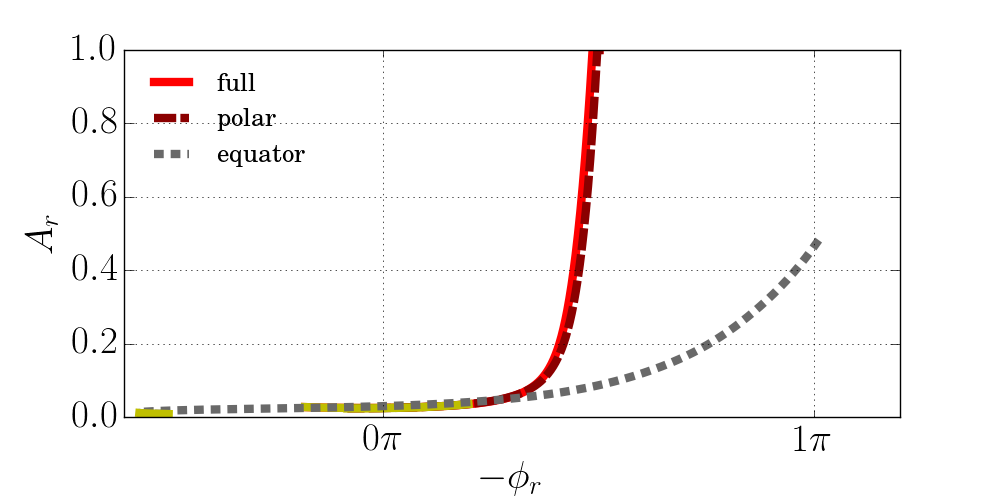}
    \includegraphics[width=\columnwidth]{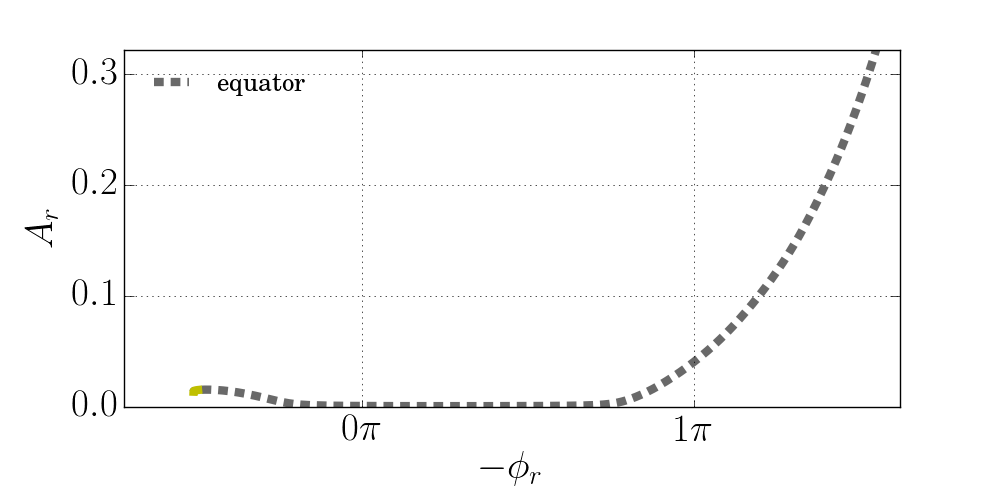}
	\caption{Amplitude versus phase variation for a magnetic field of $2$~kG. Top panel, for a mode with a frequency of $2.5$~mHz
    and a degree $l=0$, and an observer pole-on. 
    Bottom panel, for a mode with a frequency of $2.7$~mHz
    and a degree $l=0$, and an observer pole-on. 
    The solid red line shows the amplitude-phase variation for integration over the 
    full visible disk, 
    while the long dashed dark-red line shows the result for a spot around the pole 
    and the short dashed grey line the result for an equatorial belt. 
    The yellow part of the curves marks the region from the bottom of the photosphere to the bottom of the isothermal atmosphere.
We use the negative of $\phi_r$ in this figure to facilitate the comparison with the 
observational works (e.g. \citet{ryabchikova2007pulsation3D} ),
where the fit is often done to a function of the form $A_r 
\cos(\omega t -\phi_r)$, rather than the one used in our definition  (cf. eq. \ref{eq:am_pha}).}
    \label{fig:com_spots}
\end{figure} 

A common way found in the literature to analyze the observations is to combine the amplitude and phase variations 
in an amplitude-phase diagram. 
Here we perform a similar diagram based on the model results, 
for a chemical spot in the pole and a chemical spot in the equator. For simplicity, we shall consider that the chemical contrast is maximum, i.e., that only the regions inside the spot of a given element contribute to the radial velocity measured from that element. 
As we can see in Fig.\ \ref{fig:com_spots}, top panel, 
in the poles the variation in phase is smaller than in the equator, 
and also the amplitude can reach higher values.
This difference between the polar spot and the equatorial spot can be seen also in the observations. 
As an example, in the case of the element Yttrium, Y, 
that is found to be more significantly shifted from the magnetic pole than, e.g., Nd and Pr \citep{luftinger2010magnetic}, 
the radial velocities are found to have a small amplitude but to show a greater variation in phase
\citep{ryabchikova2007pulsation3D}, following the same behavior as the equator spot  in Fig. \ref{fig:com_spots}.
That is in contrast with the behavior found for other elements, concentrated in the polar regions, 
whose amplitude-phase variation behaves more like the polar spot in the same figure.
   
A claim found in several observational papers is that a node can sometimes be seen in the outer parts of the atmosphere of roAp stars. 
Based on physical grounds, in a model atmosphere like the one adopted here 
we do not expect a node anywhere in the outer atmosphere of the star.
Even for relatively low magnetic fields, the magnetic and acoustic waves are decoupled throughout most of the atmosphere.
The amplitude of the magnetic waves is constant and has a characteristic scale that is larger than the atmosphere,
thus, it cannot show a node.
Moreover, the amplitude of the acoustic waves, with an exponential growth, is either non-oscillatory, 
when the frequency is below the
critical cut-off frequency $\omega_{c}$, or has an oscillatory behavior that changes with time, 
when the frequency is above $\omega_{c}$.
Therefore, in such a model, any node detected in the outer atmosphere must be only apparent, resulting from the projection and integration
of the velocity field over the visible disk or part of it.   
We have shown an example of how that observational illusion can occur, in the case 7. 
In Fig.\ \ref{fig:com_spots}, bottom panel, we can see the amplitude versus phase variation diagram for this case. 
The node is evident in the grey dashed line, 
as we can see that the amplitude first decreases, then 
goes through a minimum of almost a pi long over a short variation of radius, and grows back again. 
This kind of behavior can be seen in the amplitude versus phases diagram of 33 Lib and 10 Aql \citep{ryabchikova2007pulsation3D,sachkov2008pulsations}.

On the other hand, true node-like features may be physically expected if sharp structural variations, 
capable of reflecting partially the acoustic waves, are present in the atmosphere. 
That kind of phenomena has been discussed in different contexts,
including in the transition between the chromosphere and corona in the sun \citep{balmforth1990mixing}
and has been found in models of roAp stars presented by
\citet{saio2010pulsations,saio2012pulsation}. 
In the latter, the authors compare the phase and amplitude variations
of models that best fit two particular roAp stars. Of particular
relevance, in the first of these studies the authors discuss the
impact on the phase and amplitude variations of
using different atmospheric models, by comparing the results
obtained with a standard Ap atmosphere, adopted from
\citet{shibahashi1985}, with those obtained with a model atmosphere that
accounts for the stratification of chemical elements observed in roAp
stars, adopted from \citet{shulyak2009model}. The authors show that
the latter model, characterized by a temperature inversion around the
atmospheric layers where Nd and Pr accumulate, provides a better
agreement with the observations, emphasizing the importance of using
an empirical, self-consistent model atmosphere, derived specifically for the star under
consideration, when attempting to perform detailed
modelling of a given star. We invite the reader to have a look at the
interesting discussion presented by these authors for further details. 

Evidence for non-standard temperature gradients, 
including temperature inversions, has been found in a number of Ap stars. 
These abnormal temperature gradients are linked to a chemical stratification of elements, 
in particular a significant accumulation of REEs in the outer atmospheric layers \cite[e.g.][]{shulyak2009model,shulyak2010realistic}. 
Moreover, possible vertical magnetic field gradients have been investigated by several studies \cite[e.g.][]{nesvacil2004probable,kudryavtsev2012magnetic,rusomarov2013three,hubrig2018magnetic}. 
The complexity of the element distribution in the atmospheres of Ap stars and the simultaneous radial and horizontal inhomogeneities,
however, render some difficulty to the interpretation of the detected magnetic field variation with height, 
leading, at times, to contradictory statements. 
For example, the presence of a radial magnetic field gradient has been corroborated by a recent study of the strongly peculiar roAp star HD 101065 (Przybylski's star) \cite{hubrig2018magnetic}. 
However, no such gradient was found for another roAp star HD 24712 \citep{rusomarov2013three}.
In any case, 
the main impact on pulsations of this complexity of the Ap stars'
atmospheric structure is expected to come from the possible sharp
temperature gradients, 
which, as mentioned above, will induce partial reflection of the acoustic waves. In particular, 
that partial reflection is likely in the origin of the quasi-nodes discussed in Saio's work.

Two stars have been argued to show a node in the atmosphere, 33 Lib and $10$ Aql
\citep{mkrtichian2003radial,elkin2008pulsation,sachkov2008pulsations}.
These two stars have a main frequency above the acoustic cut-off frequency 
and very long rotational periods.
Because of the latter the magnetic field structure and 
the surface distribution of elements cannot be derived making it difficult a direct comparison with our model. 
Nevertheless, in the light of the understanding of the problem provided by the present work we can confidently   
conclude that either we are in the presence of an apparent node,
resulting from the cancellation effect of the acoustic and magnetic waves' contributions to the integral, 
or sharp variations in the atmospheric structure of these roAp stars are capable of significantly reflecting the acoustic waves. Checking the latter possibility requires adopting a more realistic atmospheric model, which we will do in a future work. 

In conclusion, we find that the behaviour of the radial velocity in our magnetic model resembles that inferred from high-resolution spectroscopic time-series of roAp stars, 
both in what concerns the amplitude and phase variations throughout the atmosphere. 
Quantitative comparisons and further test to the model shall be carried out in a follow up work directed at the modelling of particular stars, 
in which the atmospheric structure to adopt will be one derived from empirical self-consistent modeling of the stellar spectra.

\section*{Acknowledgements}

P.Q.M. gratefully acknowledge the financial support of CONICYT/Becas Chile.
This work was also supported by the Portuguese Science foundation (FCT) through national funds (Investigador
contract of reference IF/00894/2012 and UID/FIS/04434/2013) and POPH/FSE (EC) by
FEDER through COMPETE2020 (POCI-01-0145-FEDER-030389 and POCI-01-0145-
FEDER-007672). Funds were also provided by the project FCT/CNRS: PICS 2014.
O.K. acknowledges financial support from the Knut and Alice Wallenberg Foundation, 
the Swedish Research Council, and the Swedish National Space Board.




\bibliographystyle{mnras}








\bsp	
\label{lastpage}
\end{document}